\RequirePackage{fix-cm}
\documentclass[smallextended]{svjour3}       
\smartqed  
\usepackage{graphicx}
\usepackage{url}
\newcommand{\aap}{A\&A}
\newcommand{\apj}{ApJ}
\newcommand{\mnras}{MNRAS}
\newcommand{\apss}{APSS}
\newcommand{\procspie}{Proc. SPIE}
\newcommand{\pasp}{PASP}
\newcommand{\na}{Nature Astronomy}
\newcommand{\apjl}{ApJL}
\newcommand{\aj}{AJ}
\begin{document}

\title{Grazing, non-transiting disintegrating exoplanets observed with the planned \textit{Ariel} space observatory
}
\subtitle{A case study using Kepler-1520b}

\titlerunning{Kepler-1520b in a grazing regime, observed by {\it{Ariel}}}        

\author{Zolt\'{a}n Garai        
}

\authorrunning{Z. Garai} 

\institute{Z. Garai$^{1,2,3}$ \at
              $^1$MTA - ELTE Exoplanet Research Group, 9700 Szombathely, Szent Imre h. u. 112, Hungary\\
              $^2$ELTE Gothard Astrophysical Observatory and Multidisciplinary Research Centre, 9700 Szombathely, Szent Imre h. u. 112, Hungary\\
              $^3$Astronomical Institute, Slovak Academy of Sciences, 05960 Tatransk\'a Lomnica, Slovakia\\
              \email{zgarai@gothard.hu}           
}

\date{Received: date / Accepted: date}

\maketitle

\begin{abstract}
Disintegrating/evaporating rocky exoplanets can be observed not only as transiting planets, but also in a grazing, non-transiting regime, where the solid body of the planet does not transit, but part of the comet-like tail can transit. In this case the forward scattering on the escaping particles is the dominant process, which amplifies the photometric signal of the parent star detected by the observer. The change in the flux is small, about $10^{-3}$ (1000 ppm) at the best properties of the planetary system, but if the observation is enough precise, the detection is possible. The planned \textit{Ariel} space observatory is designed to achieve a stability of $< 100$ ppm (the goal is 10 ppm) over the temporal bandwidth of the transit, typically less than 10 hours. In this case study we took the disintegrating exoplanet Kepler-1520b and changed the orbital properties of the system to get a grazing, non-transiting orbit scenario, and investigated, how different particle sizes, species, \textit{Ariel} observational channels, and other factors affect the amplitude of the forward-scattering peak, and the detectability of the scattering event. Our most important result is that the forward-scattering amplitude is not sensitive to the dust composition, but is very sensitive to the particle size, observational channel, and other factors. These factors can reduce mainly the detectability of 1-micron grains. 0.1-micron grains will be detectable at short wavelengths. 0.01-micron grains generate long and very small forward scattering amplitude, which is below the detection limit. Based on our results we can conclude that using \textit{Ariel} will be possible to detect and investigate not only transiting, but also grazing, non-transiting disintegrating exoplanets based on the forward scattering. From the viewpoint of such objects the big advantage of \textit{Ariel} will be the possibility of multiwavelength observations. 

\end{abstract}

\section{Introduction}
\label{intro}

\begin{figure*}
\centering
\centerline{
\includegraphics[width=60mm]{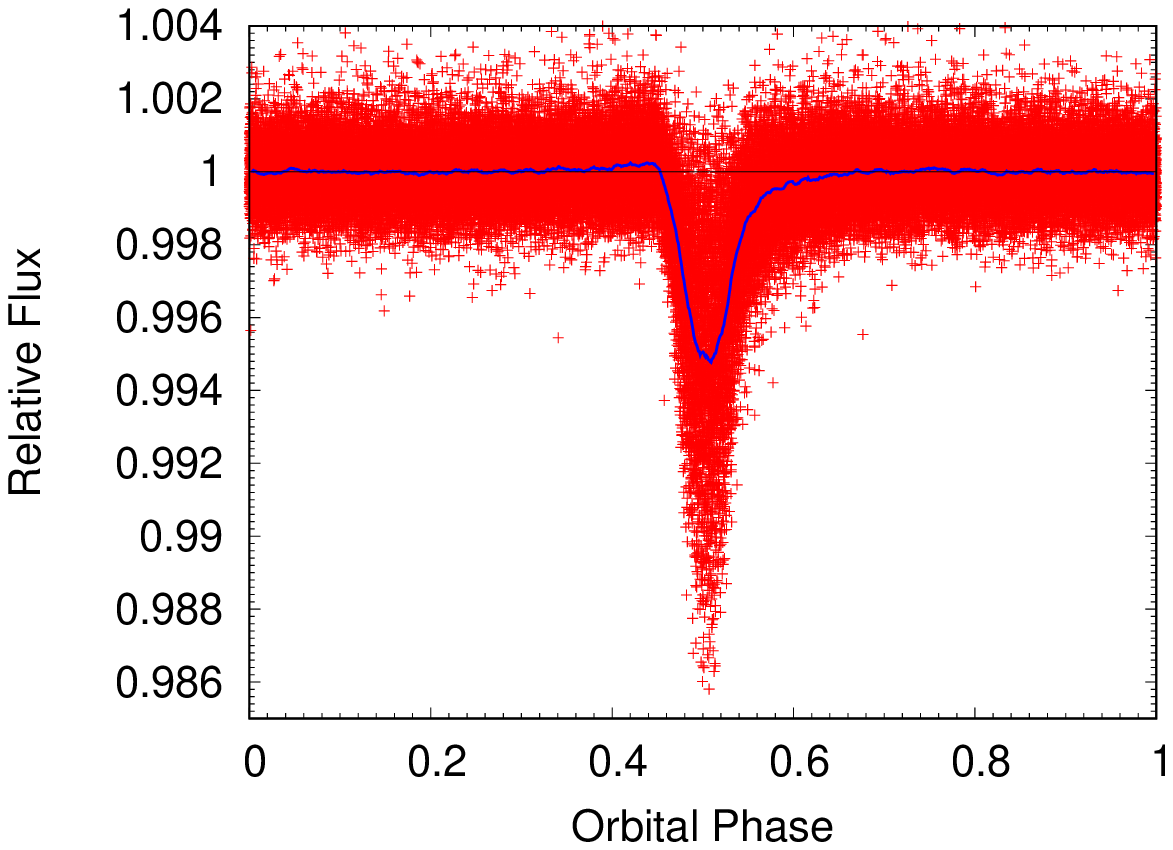} 
\includegraphics[width=60mm]{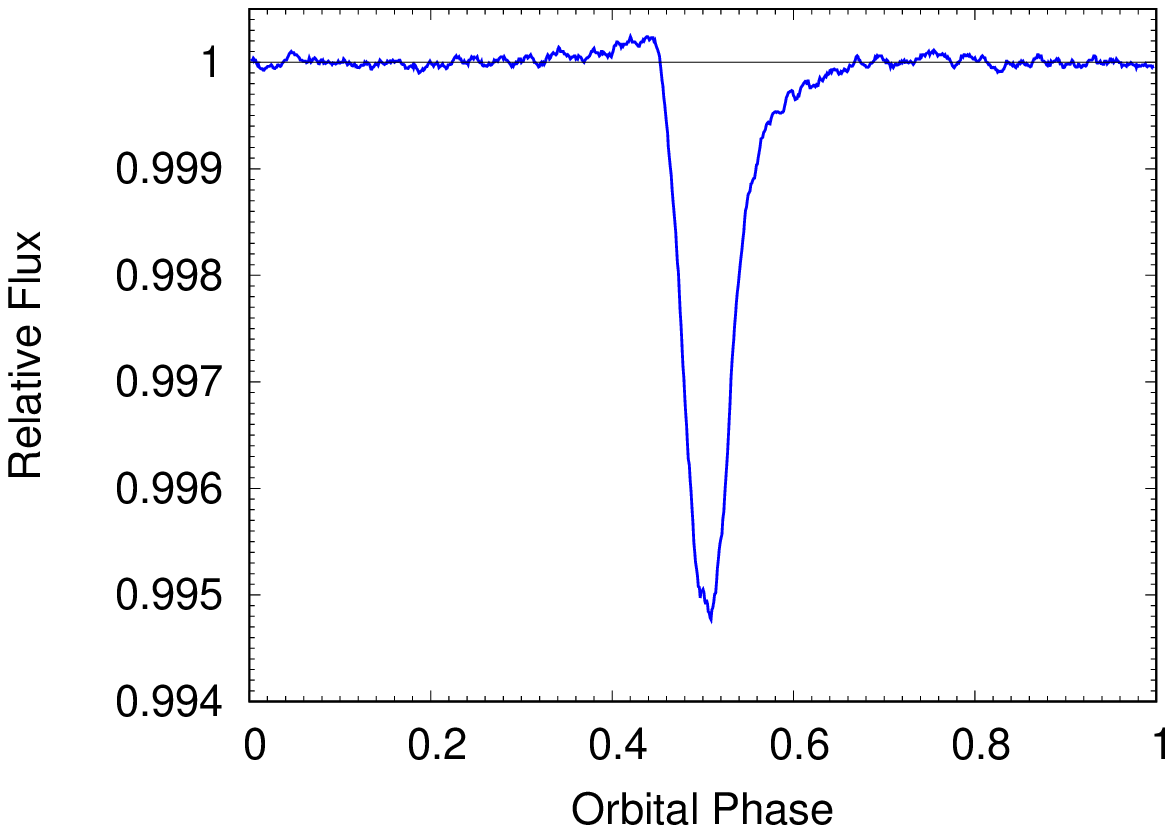}} 
\caption{Phase-folded and averaged transit light curve of Kepler-1520b, smoothed using the running window technique. The window width and step was 0.01 and 0.001, respectively (in units of phase). Red
points are \textit{Kepler} observations and the blue line represents averaged data (left-hand panel). The same averaged data, focused to better visualize the asymmetry of the light curve (right-hand panel).}
\label{Cele}    
\end{figure*}

Three exoplanets have been discovered with the \textit{Kepler} mission \cite{Borucki1} that are inferred to have tails of dusty effluents trailing behind, or ahead of them in orbit about their host star. The first exoplanet with a comet-like tail, Kepler-1520b, was discovered by \cite{Rappaport1}, and was found to be a close-in exoplanet with an extremely short orbital period of $P_\mathrm{orb} = 0.65356(1)$ d. The host star that is apparently being occulted is Kepler-1520, a $V = 16.7$ mag K-dwarf with a $T_\mathrm{eff} = 4677(82)$ K. The shape of the transit is highly asymmetric (see Fig. \ref{Cele}). It shows a significant brightening just before the eclipse -- pre-transit brightening, sharp ingress followed by a short sharp egress and long smooth egress, and a weak post-transit brightening. Moreover, the planet exhibits strong variability in the transit core on the timescale of one day \cite{Rappaport1,Bochinski1}, and a variability in the egress on the timescale of about 1.3 years \cite{Budaj1}. \cite{Rappaport1} suggested that the planet's size is not larger than Mercury, and is slowly disintegrating/evaporating, creating a comet-like tail. \cite{Brogi1} and \cite{Budaj1} first validated the disintegrating-planet scenario using a model and both found that dust particles in the tail have typical radii of about 0.1 - 1 micron. Both brightenings are caused by the forward scattering on dust particles in the tail. Strong variability in the transit depth is a consequence of changes in the cloud optical depth. \cite{Perez1} proposed a model of the atmospheric escape via the thermal wind that is only effective for planets, which are less massive than Mercury. Gravity of more massive planets would provide too deep potential barrier for the wind. Later, two more exoplanets were discovered, KOI-2700b and K2-22b, whose transit shapes show evidence of a similar comet-like tail \cite{Rappaport2,Sanchis1}.     

The ongoing \textit{Transiting Exoplanet Survey Satellite} (\textit{TESS}) mission \cite{Ricker1} seems to be also promising to increase the number of such objects. Recently, three distinct dipping events in the light curve of $\beta$ Pictoris were identified based on the \textit{TESS} observations by \cite{Zieba1}. The dips are asymmetric in nature and are consistent with a model of an evaporating comet with an extended tail crossing the disc of the star. \textit{TESS} studying more stars than were targeted by \textit{Kepler} and \textit{K2}. Therefore, we can expect more similar discoveries in the near future. Moreover, the host stars are likely to be brighter and therefore easier to do follow-up studies from the ground. 

The planned \textit{Ariel} space mission \cite{DaDeppo1,Pascale1,Tinetti1}, expected to be launched in 2029, has also potential to be successful in detection of disintegrating rocky exoplanets, or exocomets. During its 3.5-years operations from L2 orbit \textit{Ariel} will continuously observe exoplanets transiting their host star. It is designed to achieve a stability of $< 100$ ppm (the goal is 10 ppm) over the temporal bandwidth of the transit, typically less than 10 hours. Moreover, the observations will be obtained in several channels, centered at 0.55, 0.70, 0.95, 1.25, and 1.65 microns. The optical properties of dust grains in the tail (hence also the transit depth) may vary with the wavelength. Small grains scatter light in the Rayleigh scattering regime, which decreases steeply with the wavelength. On the other hand, large particles attenuate the light in the geometrical optics regime, which does not depend on the wavelength, see e.g., Fig. 13 in \cite{Croll1}. The optical properties of dust depend mainly on the particle size and their chemical composition. Consequently, multiwavelength transit observations can put constraints on these two parameters. 

Disintegrating exoplanets discovered by the \textit{Kepler} space telescope have host stars too faint to be targeted in the \textit{Ariel} core survey, but nearby analogues of such disintegrating exoplanets orbiting bright stars must exist. These bright stars would be attractive targets for \textit{Ariel} core survey. We propose the target selection based on the results of the Dispersed Matter Planet Project \cite{Haswell1}. This project was partly motivated by the aim of finding nearby disintegrating exoplanets and their progenitors. Some highly irradiated close-in exoplanets orbit stars showing anomalously low chromospheric emission. The authors in \cite{Haswell1} attribute this deficit to absorption by circumstellar material replenished by mass loss from disintegrating planets. Hence, anomalously low chromospheric emission can indicate disintegrating planets. The first possible transiting disintegrating exoplanet in the framework of this project was already reported in the $V = 7.98$ mag system DMPP-1 (HD38677) by \cite{Jones1}. Their results also suggest that disintegrating rocky exoplanets can co-exist with hot/warm giant planets.

There is one more aspect of disintegrating exoplanets, and this is the advantage that there is more phase space available to detect some of these objects, even if the planet does not transit the host star, i.e., via the forward-scattering signal alone. In the transiting case the extinction by dust during the transit removes more light from the beam than is scattered into it. Thus, the forward-scattering component of the light is best seen either just prior to ingress, or just after egress, but with reduced amplitude over the larger peak that is obscured by the transit. The highest amplitude portion of the forward-scattering peak is obscured by the transit part of the light curve, where extinction nullifies the effect of the forward scattering. This raises the possibility that grazing, non-transiting exoplanets with dusty tails, where the solid body of the planet does not transit, but part of the comet-like tail can transit, might have a distinct and detectable signature via this same forward-scattering peak, as it was pointed out by \cite{DeVore1}. The authors calculated that the amplitude of the scattering peak could be in the range from 0.00005 to 0.0005 (from 50 to 500 ppm), depending on the orbit inclination angle. Contrary to \cite{DeVore1}, in our case study we took the disintegrating exoplanet Kepler-1520b and changed the orbital properties of the system to get a grazing, non-transiting orbit scenario, and investigated, how different particle radii, species, \textit{Ariel} observational channels, and other factors affect the amplitude of the forward-scattering peak, and the detectability of the scattering event. 

The paper is organized as follows. We first calculated the optical properties of the dust used in our analysis. This is briefly described in Section \ref{optprop}. We then modeled the phase-folded and averaged transit light curve of Kepler-1520b using the {\tt{Shellspec}} code. This procedure is detailed in Section \ref{transitmod}. Section \ref{grazingmod} is the main part of our work. We took the orbital and planetary properties of Kepler-1520b discussed in Section \ref{transitmod}, changed only the orbital inclination of the system to get a grazing, non-transiting orbit scenario, and performed several modeling using the {\tt{Shellspec}} code. In Section \ref{discuss} we discussed three important factors, which could affect the results described in Section \ref{grazingmod}. Finally, our findings are summarized in Section \ref{conc}.  

\section{Calculation of the optical properties of the dust}
\label{optprop}

Dust can absorb the impinging radiation and convert it directly into heating of the grains. This process is called \textit{absorption}, or \textit{true absorption} and it is quantified by the \textit{absorption opacity}. Dust can also scatter radiation in a process called \textit{scattering} without being heated. This process is quantified by the \textit{scattering opacity}. The sum of the absorption opacity and the scattering opacity is the \textit{total opacity} (or simply the \textit{opacity}). Furthermore, scattering can be highly asymmetric, a property that is described by means of the phase function, which depends on the scattering angle (the deflection angle from the original direction of the impinging radiation). The most prominent feature is a strong forward scattering, when the scattering angle is nearly zero.

Optical properties of selected species were calculated by \cite{Budaj2}. The tables, which contains phase functions, opacities, albedos, equilibrium temperatures, and radiative accelerations of dust grains in exoplanet systems are freely available on the web\footnote{See \url{https://www.ta3.sk/~budaj/dust/deirm}}. The tables cover the wavelength range from 0.2 to 500 microns and 21 particle radii from 0.01 to 100 microns for several species. Their assumptions include spherical grain shape, Deirmendjian particle size distribution, and Mie theory. From these tables we selected opacities and phase functions for alumina, enstatite, forsterite, olivine with 50\% magnesium, pyroxene with 40\% magnesium, and iron, for the particle radii of 0.01, 0.1, and 1 micron. We also have to take into account the fact that from the viewpoint of dust particles the parent star has a non-negligible angular dimension on the sky. Dust particles of Kepler-1520b in the distance of $a = 2.77~\mathrm{R}_{\odot}$ from the parent star view this star as a disk with an angular diameter of about $26^{\circ}$. The true dimension of this star is $R_\mathrm{s} = 0.66~\mathrm{R}_{\odot}$ \cite{Rappaport1}. To take this effect into account we have to split the stellar disk into elementary surfaces and integrate the phase function over the disk. We calculated phase functions with a very fine step in the interval from 0 to $13^{\circ}$, because of the strong forward scattering and, consequently, the disk averaged phase function with a very fine step near the edge of the stellar disk. For this purpose we applied the software {\tt{Diskaver}}\footnote{See \url{https://www.ta3.sk/~budaj/dust/deirm/diskaver}}. It assumes a quadratic limb darkening of the stellar surface:  

\begin{equation}
\label{qadraticlimbdarklaw}
I_\nu = I_\nu(0)[1-u_1(1 - \cos \theta) - u_2(1 - \cos \theta)^2],
\end{equation}

\noindent{where $I_\nu(0)$ is the intensity perpendicular to the surface of the source and $\theta$ is the angle between the line of sight and a normal to the surface. The quadratic limb darkening coefficients $u_1$ and $u_2$ were linearly interpolated based on the stellar parameters of $T_\mathrm{eff} = 4677~\mathrm{K}$, $\log g = 4.60$ (cgs), and $\mathrm{Fe/H}=-0.18$ \cite{Rappaport1} for the passbands $V$, $i$, $z$, $J$, and $H$, which correspond to the \textit{Ariel} observational channels of 0.55, 0.70, 0.95, 1.25, and 1.65 microns, respectively. Subsequently, we also calculated the quadratic limb darkening coefficients for the \textit{Kepler} passband. During this step we used the on-line applet {\tt{EXOFAST - Quadratic Limb Darkening}}\footnote{See \url{http://astroutils.astronomy.ohio-state.edu/exofast/limbdark.shtml}}, which is based on the IDL-routine {\tt{QUADLD}} \cite{Eastman1}. This software interpolates the \cite{Claret1} quadratic limb darkening tables. Calculations with the software {\tt{Diskaver}} were performed on the wavelengths of 0.55, 0.60, 0.70, 0.95, 1.25, and 1.65 microns for consistency with the \textit{Ariel} and \textit{Kepler} passbands.}

\section{The transit model of Kepler-1520b}
\label{transitmod}

We first modeled the phase-folded and averaged transit light curve of Kepler-1520b (see Fig. \ref{Cele}, right-hand panel). This analysis was previously carried out by \cite{Budaj1} using the {\tt{Shellspec}} code \cite{Budaj3}, but because we used other models of the comet-like tail, we had to repeat this analysis. During our analysis we also used the same code\footnote{See \url{https://www.ta3.sk/~budaj/shellspec.html}}, but its version No. 39. This code calculates the light curves and spectra of interacting binaries or exoplanets immersed in the three-dimensional (3D) circum-stellar, or circum-planetary environment. It solves simple radiative transfer along the line of sight and the scattered light is taken into account under the assumption that the medium is optically thin. A number of optional objects (such as a spot, disk, stream, ring, jet, shell) can be defined within the model, or it is possible to load a precalculated model from an extra file. Synthetic light curves or trailing spectrograms can be produced by changing our viewpoints on the 3D object.

For our purpose we used an optional object in the form of a ring, which we subsequently modified. We can model a comet-like tail as part of a ring with a non-negligible thickness around a central star. During our calculations we assumed a spherical and limb-darkened central star with a radius of $R_\mathrm{s} = 0.66~\mathrm{R}_{\odot}$, mass of $M_\mathrm{s} = 0.76~\mathrm{M}_\odot$, and effective temperature of $T_\mathrm{eff} = 4677~\mathrm{K}$ \cite{Rappaport1}, located in the geometrical center of the ring. We modeled the comet-like tail as part of a ring with a radius of $a = 2.77~\mathrm{R}_{\odot}$. Its geometrical cross-section is monotonically enlarging from the planet to the end of the ring, which is located at $60^{\circ}$ behind the planet. At this point the ring is truncated. The cross-section of the ring \textit{C} and dust density along the ring $\rho$ are allowed to change with the angle \textit{t} [rad]:

\begin{equation}
\label{densitychangeA2}
\rho(t) = \rho(0)\frac{C(0)}{C(t)}e^{(|t-t(0)|A2)/\pi},
\end{equation}

\noindent{where $\rho(0)$, $C(0),$ and $t(0)$ are the dust density, cross-section, and phase angle of view at the beginning of the ring, respectively, and $A2$ is the density exponent to model the dust destruction in the tail. We can see that there is a strong degeneracy between the cross-section and the dust density at a certain phase angle. In general, if we increase the cross-section and we want to obtain an appropriate model of the observed light curve, the result is that we need to decrease the dust density. There is a degeneracy relation $C\rho=$ const. Since the dimension of the dust tail of Kepler-1520b is unknown, we arbitrarily defined a geometrical cross-section of the tail, which was preferable for our computation process (in terms of the grid density, grid dimension, and computing time). Therefore, in our calculations we assumed a dust tail with a cross-section of $0.05 \times 0.05~\mathrm{R}_\odot$ at the beginning and $0.09 \times 0.09~\mathrm{R}_\odot$ at its end. These values are also in agreement with the escape velocity from a Mercury-sized small planet (a few km.s$^{-1}$). We note that this is the first difference in comparison with \cite{Budaj1}, who used a dust-tail model with a cross-section of $0.01 \times 0.01~\mathrm{R}_\odot$ at the beginning and $0.09 \times 0.09~\mathrm{R}_\odot$ at its end.}

In the {\tt{Shellspec}} code the central star with the defined object is located in a 3D grid. The code enables the user to look on the grid from different points of view and to calculate the corresponding flux. The flux is always calculated in the observer’s line of sight. The orbit inclination angle $i$ corresponds the inclination of the intrinsic rotation axis of the model to the line of sight. At each point of view we calculated the final flux as $(s+r)/s = f$, where $s$ means modeling the flux from the parent star, $s+r$ means modeling the parent star with the ring, and $f$ is the final and normalized flux from the system. In this way we also eliminated fluctuations due to the grid structure of objects in our model. The synthetic light curves were subsequently convolved with a box-car with a width of 30 min, simulating the integration time of the \textit{Kepler} long cadence exposure. Convolved light curves were used for comparison with the observed light curve. 

\begin{figure*}
\centering
\centerline{
\includegraphics[width=60mm]{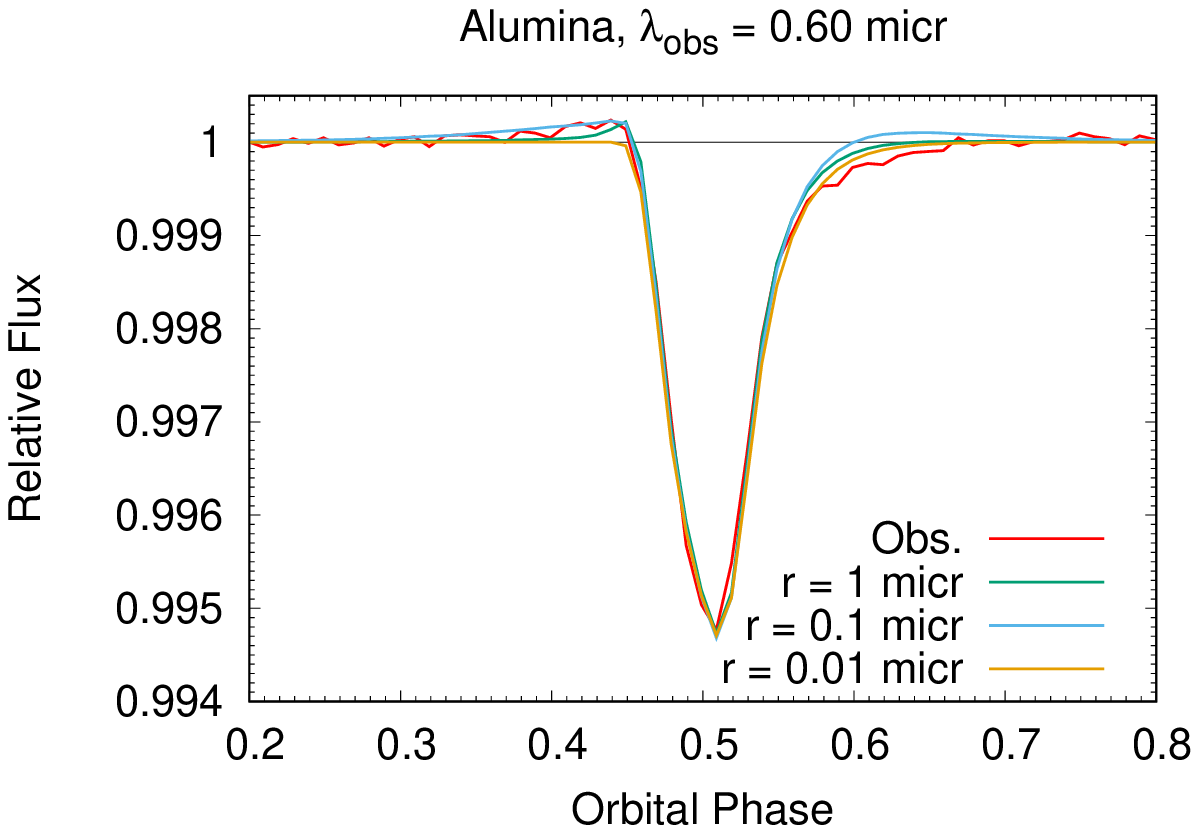} 
\includegraphics[width=60mm]{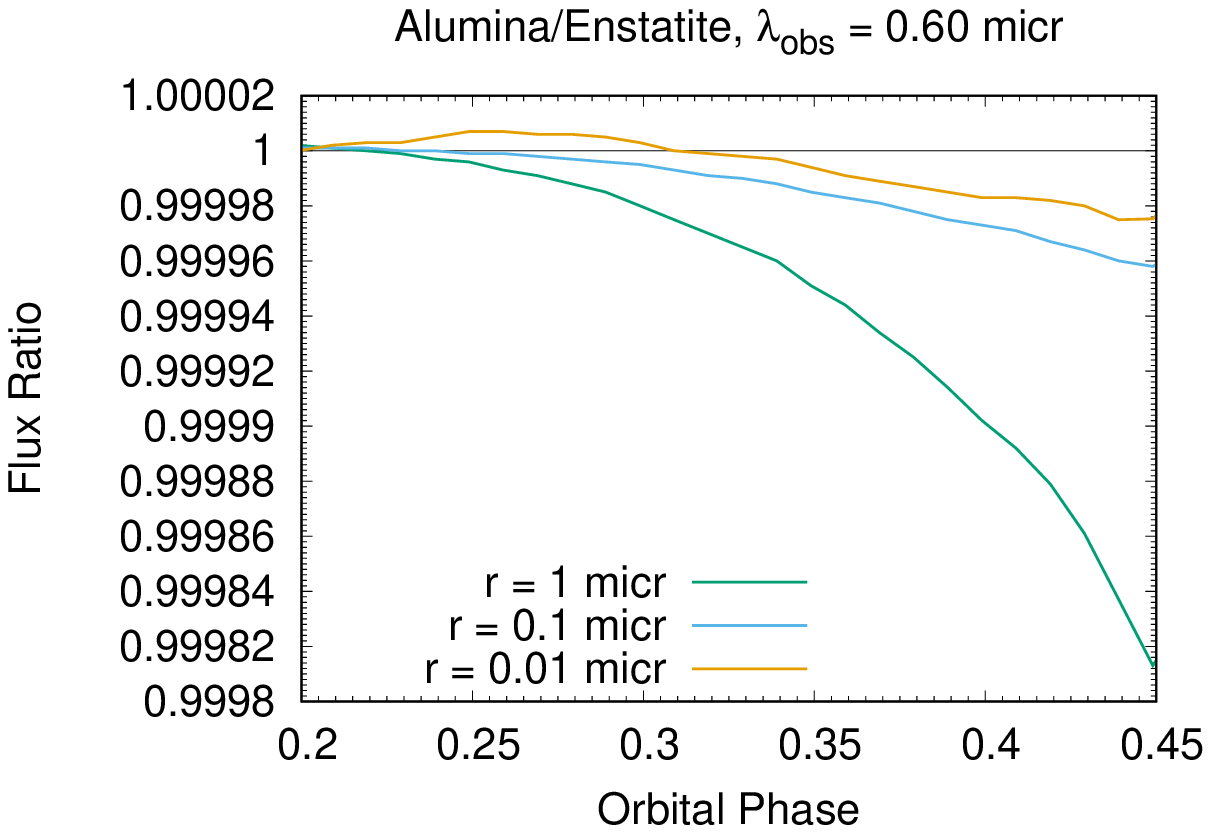}} 
\centerline{
\includegraphics[width=60mm]{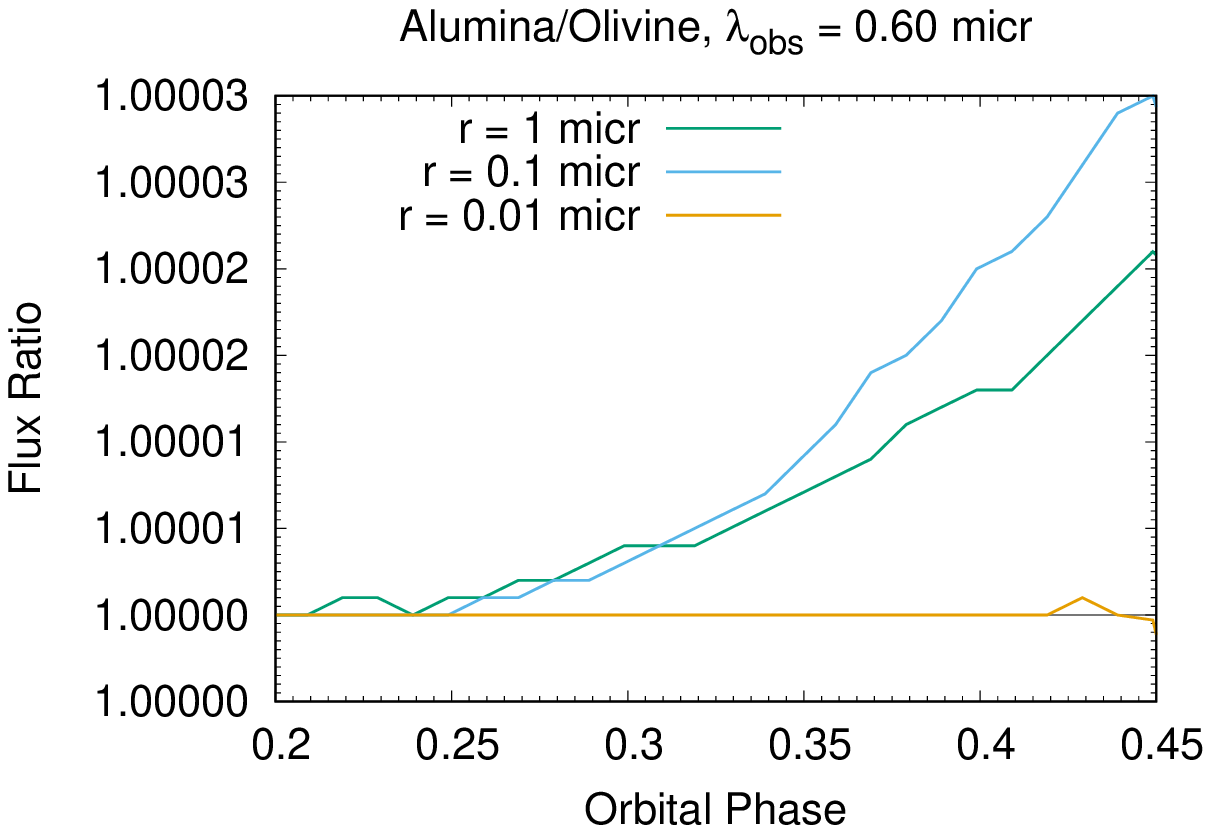} 
\includegraphics[width=60mm]{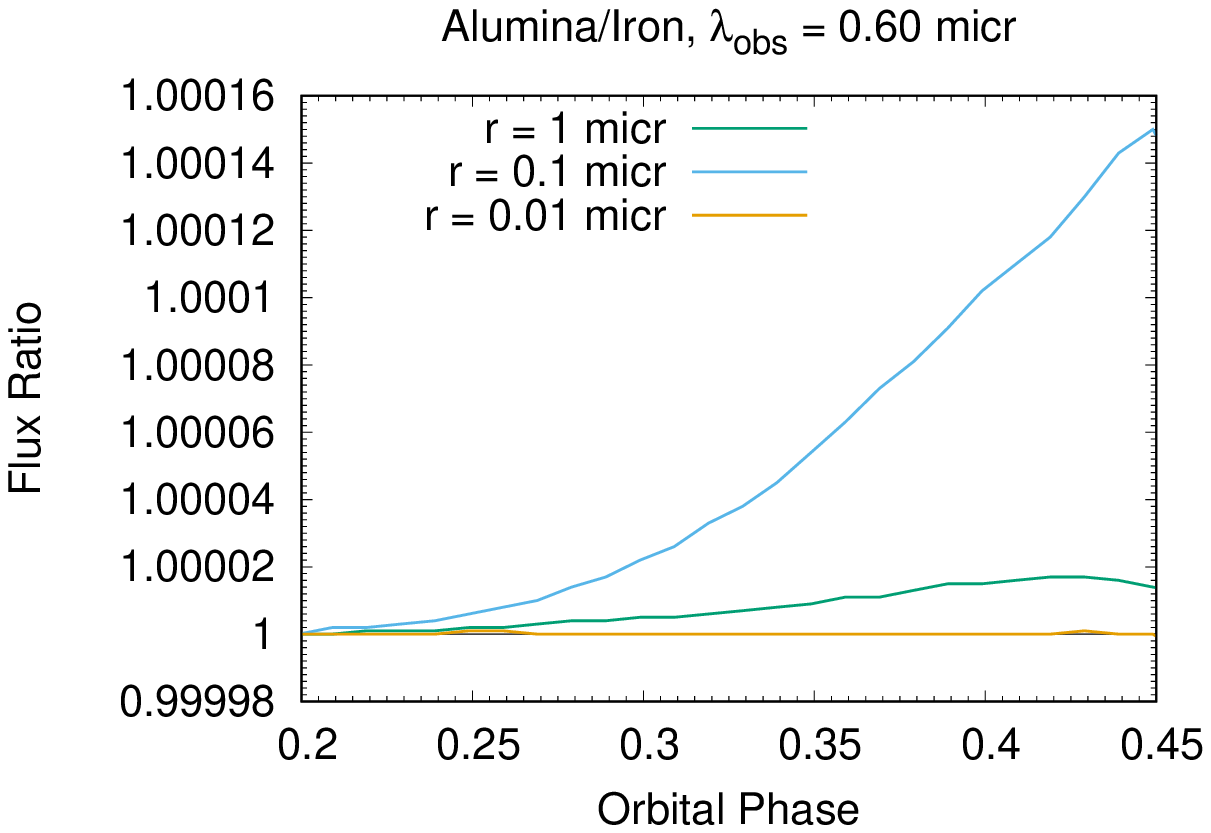}} 
\caption{Model light curves calculated for 0.01-micron, 0.1-micron and 1-micron grains of alumina, compared with the observed phase-folded and averaged transit light curve of Kepler-1520b are depicted on top left-hand panel. Flux ratios between alumina and enstatite/forsterite, alumina and olivine/pyroxene, alumina and iron calculated for the orbital phases, where the pre-transit brightening is visible, are depicted on top right-hand panel, bottom left-hand panel and bottom right-hand panel, respectively.}
\label{transitmodel}    
\end{figure*}

For the modeling process we used an iterative procedure, which applied the {\tt{Shellspec}} code as a subroutine, and which searched for the best fit, similarly as per \cite{Garai1}. Only one free parameter was adjusted during the fitting procedure, i.e., the dust density at the beginning of the ring $\rho(0)$ [g.cm$^{-3}$]. The orbit inclination angle $i$ [$^{\circ}$] and the density exponent $A2$ were fixed during the fitting procedure using the values found by \cite{Budaj1}, i.e., $i = 82^{\circ}$ and $A2 = -20$. One more free parameter -- the transit midpoint phase shift of the synthetic light curve with respect to the observed light curve ($\Delta\varphi_0$) -- was adjusted only before the modeling process and then was kept fixed to its best value. This parameter reflects the unknown mid-transit time of the planet. Every synthetic light curve was shifted in phase by $\Delta\varphi_0=-0.26$. The advantage of this treatment is that it saves computing time. A formal, quantitative goodness-of-fit was measured via determination of $\chi^2$. The best fit corresponds to the minimum value of $\chi^2$. The calculations were performed on the wavelength of 0.6 micron for consistency with the \textit{Kepler} passband. We executed 18 fitting procedures -- one for each combination of species and dust particle sizes. We note that this is the second difference in comparison with \cite{Budaj1}, who used only four species (pyroxene with 40 \% magnesium, enstatite, forsterite, and iron) and three particle sizes (0.01, 0.1, and 1 micron). The best-fitting $\rho(0)$ values are summarized in the Table \ref{rhovalues}. The observed phase-folded and averaged transit light curve of Kepler-1520b, overplotted with the best-fitting models of alumina is depicted, as an example, in Fig \ref{transitmodel}, top left-hand panel. For better effect visibility, selected flux ratios between two species are also shown on the further panels of the same figure. We note that the models composed from enstatite are very similar to the forsterite-models, also the olivine-models are very similar to the models composed from pyroxene. Contrary to \cite{Garai1}, in this case we did not derive uncertainties in the fitted $\rho(0)$ values, because it takes a lot of computing time and the scientific goal of this part of the analysis is not to investigate the light curve of Kepler-1520b, which was already performed by several authors, see e.g., \cite{Brogi1,Budaj1}, rather to prepare the next step, which is modeling Kepler-1520b as a grazing, non-transiting planet. The second simplification made during this step was that we did not include the solid body of the planet in the model. This is justified, because the planet's solid body, contrary to KOI 2700b, is negligible in comparison with the comet-like tale itself.  

Although our scientific goal was not to study the transit light curve of Kepler-1520b, based on the obtained transit models we can conclude that (1) the modified ring-model of the tail fits the observations well, except the end of the transit egress, where the models appear to over-predict this part of the light curve. This is very probably due to the dominant particles located at the end of the tail. These can have smaller radii than 0.01 micron. Fortunately, this feature does not affect our further conclusions. (2) The applied orbit inclination angle $i$ and the density exponent $A2$, derived by \cite{Budaj1}, seem to be appropriate. (3) Based on modeling of the transit we cannot determine the chemical composition of the dust ejected by Kepler-1520b. (4) Based on modeling of the forward-scattering amplitude we can conclude that the typical particle at the beginning of the tail, where is the largest concentration of the evaporating material, is the 0.1-micron grain, if the particles are composed from alumina, enstatite, forsterite, olivine with 50 \% magnesium, or from pyroxene with 40 \% magnesium. If the evaporating material is composed from alumina, olivine, pyroxene, or from iron, then 1-micron grains are also possible. Models applying 0.01-micron grains do not satisfy the observed forward-scattering amplitude. This is in agreement with the results about the typical particle size obtained by \cite{Brogi1}, \cite{Budaj1}, and by \cite{Bochinski1}.

\begin{table*}
\caption{The best-fitting $\rho(0)$ values as a result of the fitting procedure, where we modeled the observed phase-folded and averaged transit light curve of Kepler-1520b. These values were derived based on a tail-model with a cross-section of $0.05 \times 0.05~\mathrm{R}_\odot$ at the beginning and $0.09 \times 0.09\mathrm{R}_\odot$ at its end. $^{1}$With 50\% magnesium. $^{2}$With 40\% magnesium.}
\label{rhovalues}
\begin{tabular}{l|llllll}
\hline\noalign{\smallskip}
Particle size   & Alumina 	& Enstatite 	& Forsterite 	& Olivine$^1$		& Pyroxene$^2$		 	& Iron\\
$r$ [micron]    & \multicolumn{6}{c}{$\rho(0)$ ($\times 10^{-15}$) [g.cm$^{-3}$]}\\
\noalign{\smallskip}\hline\noalign{\smallskip}
0.01 	        & 14.0		& 515.0		& 410.5		& 8.1			& 17.5				& 3.35\\
0.1  		& 1.175		& 1.35		& 1.14		& 0.81			& 0.925				& 1.45\\
1  		& 9.70		& 11.0		& 10.4 		& 10.0			& 10.25				& 26.0\\
\noalign{\smallskip}\hline
\end{tabular}
\end{table*}

\section{Kepler-1520b as a grazing, non-transiting exoplanet}
\label{grazingmod}

Brightenings on the light curve of Kepler-1520b are also caused by the forward scattering on dust particles in the tail \cite{Brogi1,Budaj1}. The most dominant out-of-transit part light-curve feature of Kepler-1520b is the pre-transit brightening, (see Fig. \ref{Cele}, right-hand panel), which has the highest amplitude portion just before the transit. During the transit, extinction nullifies the effect of the forward scattering, therefore it is not visible in these orbital phases. After the transit event the forward scattering becomes again visible during the less dominant post-transit brightening. We can therefore conclude that without transits the forward-scattering peak would be the most dominant feature of the Kepler-1520b light curve. This means that theoretically it is possible to detect grazing, non-transiting disintegrating exoplanets only via this signature, as it was pointed out by \cite{DeVore1}. The question is, how different particle radii, species, observational channels, and other factors affect the amplitude of the forward-scattering peak, and the detectability of the scattering event.

\begin{figure*}
\centering
\centerline{
\includegraphics[width=60mm]{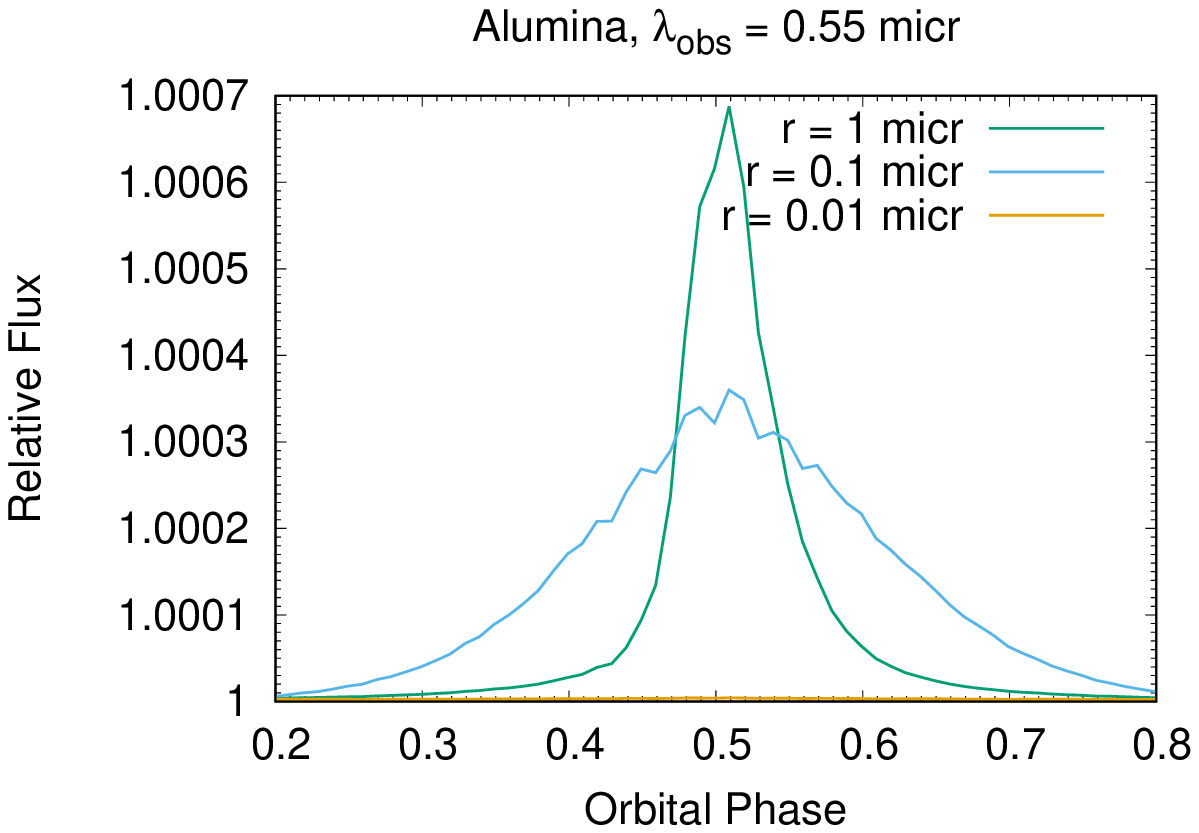} 
\includegraphics[width=60mm]{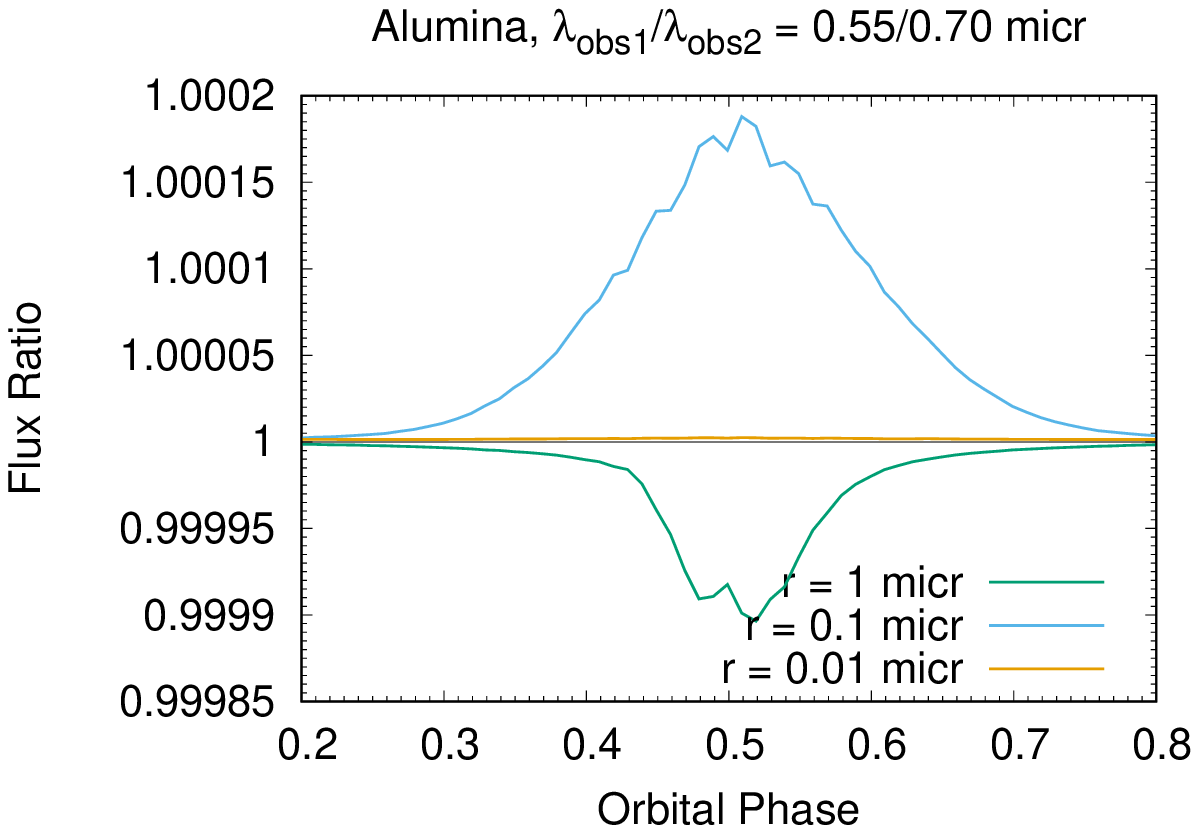}} 
\centerline{
\includegraphics[width=60mm]{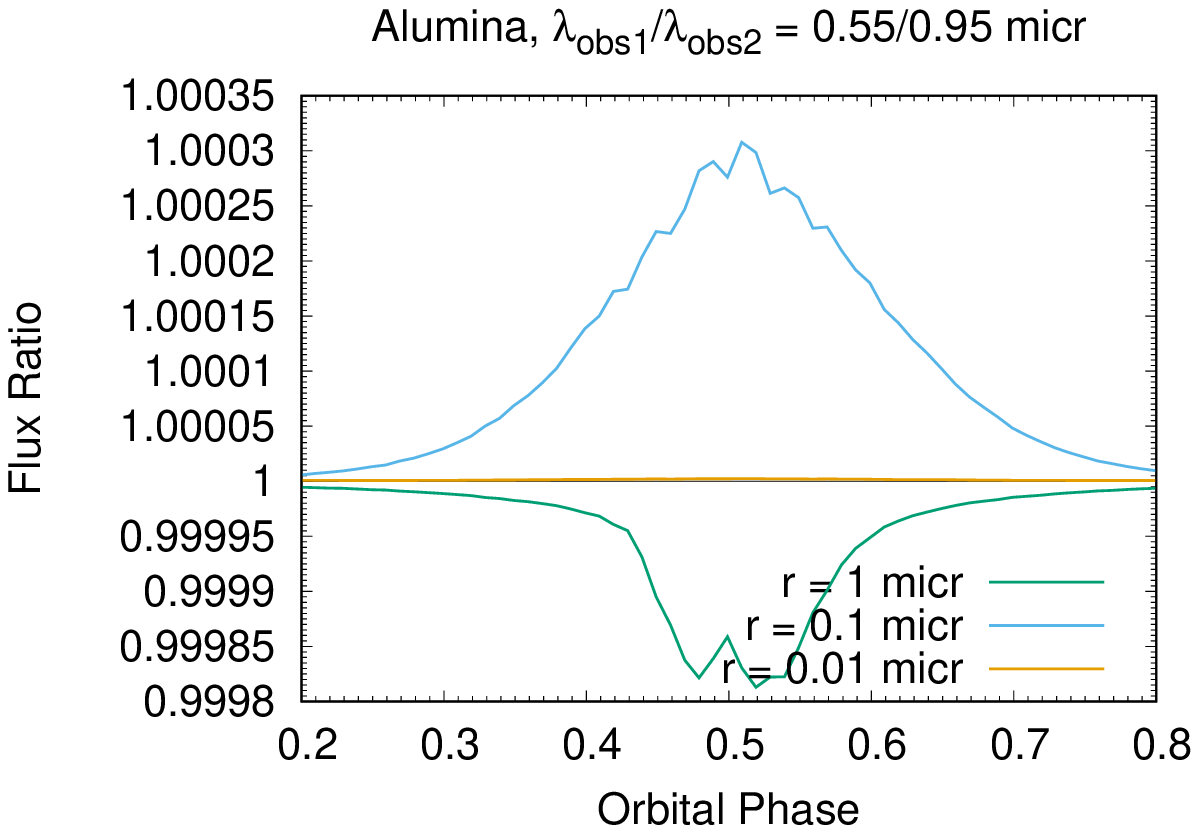} 
\includegraphics[width=60mm]{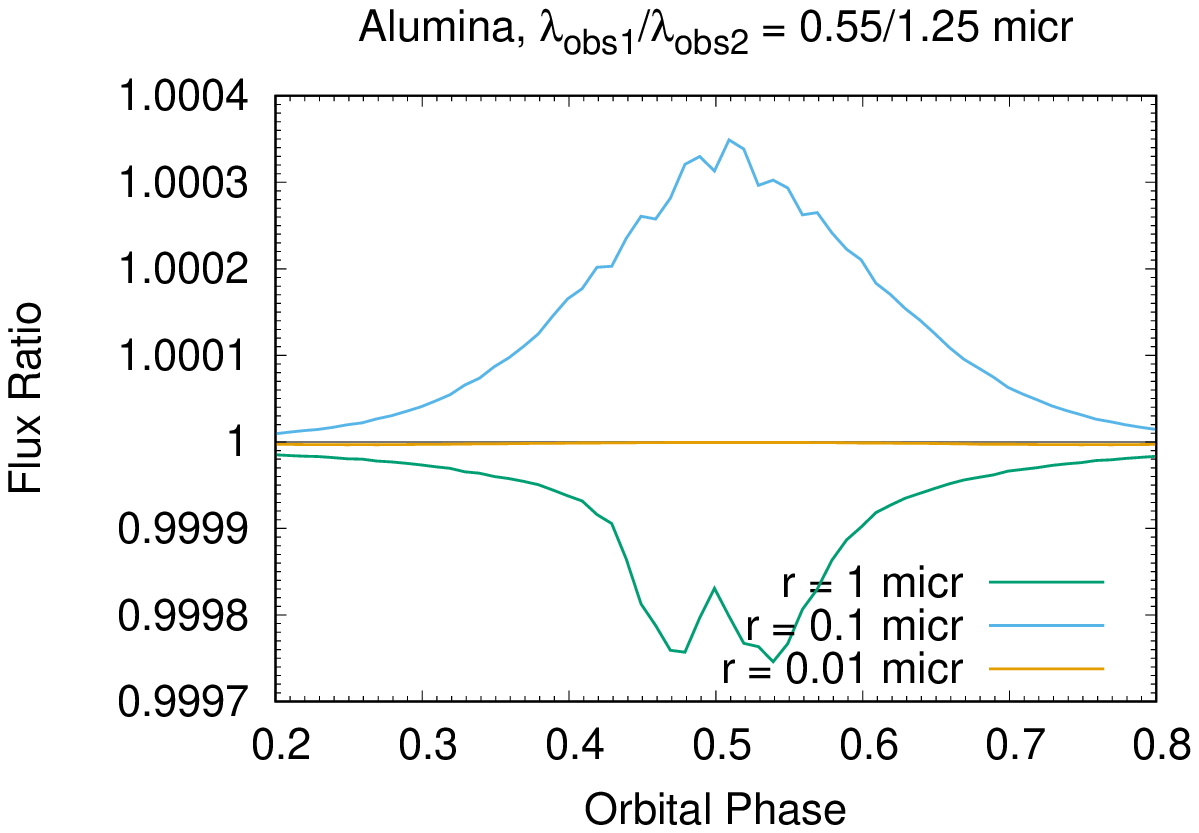}} 
\centerline{
\includegraphics[width=60mm]{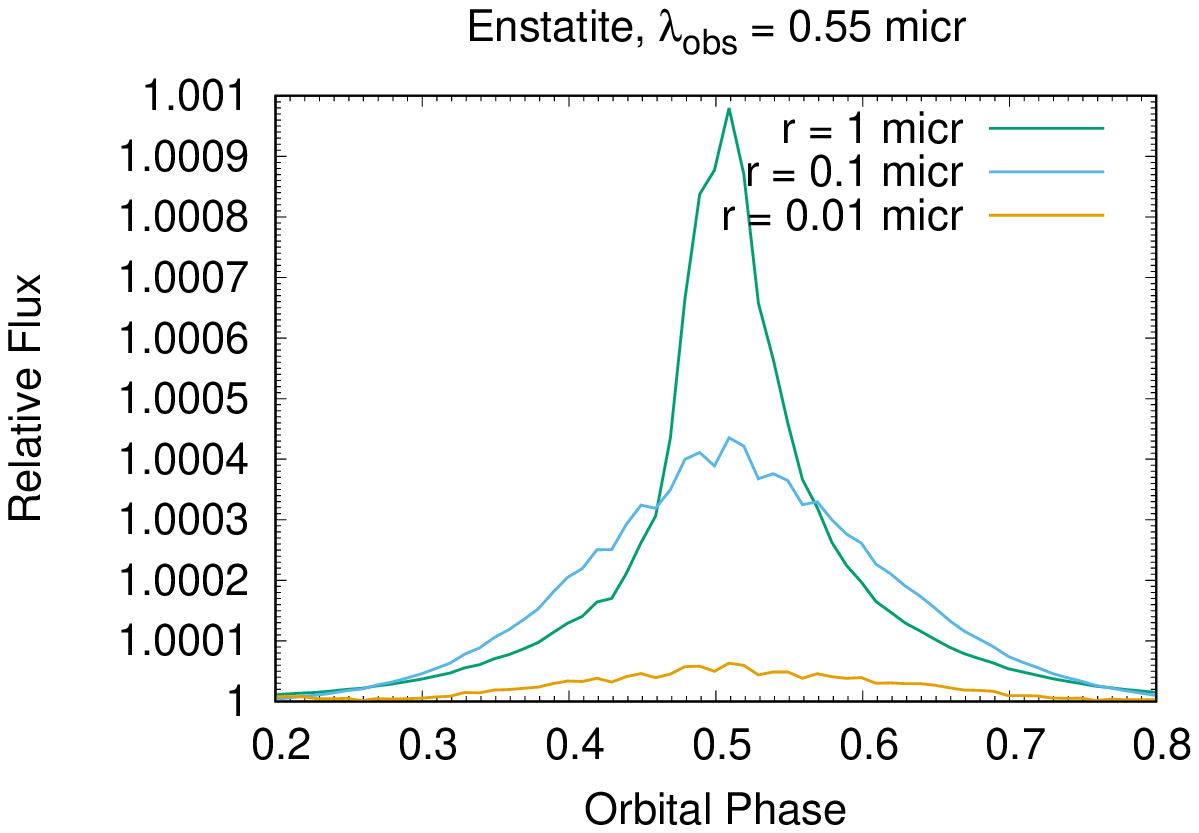} 
\includegraphics[width=60mm]{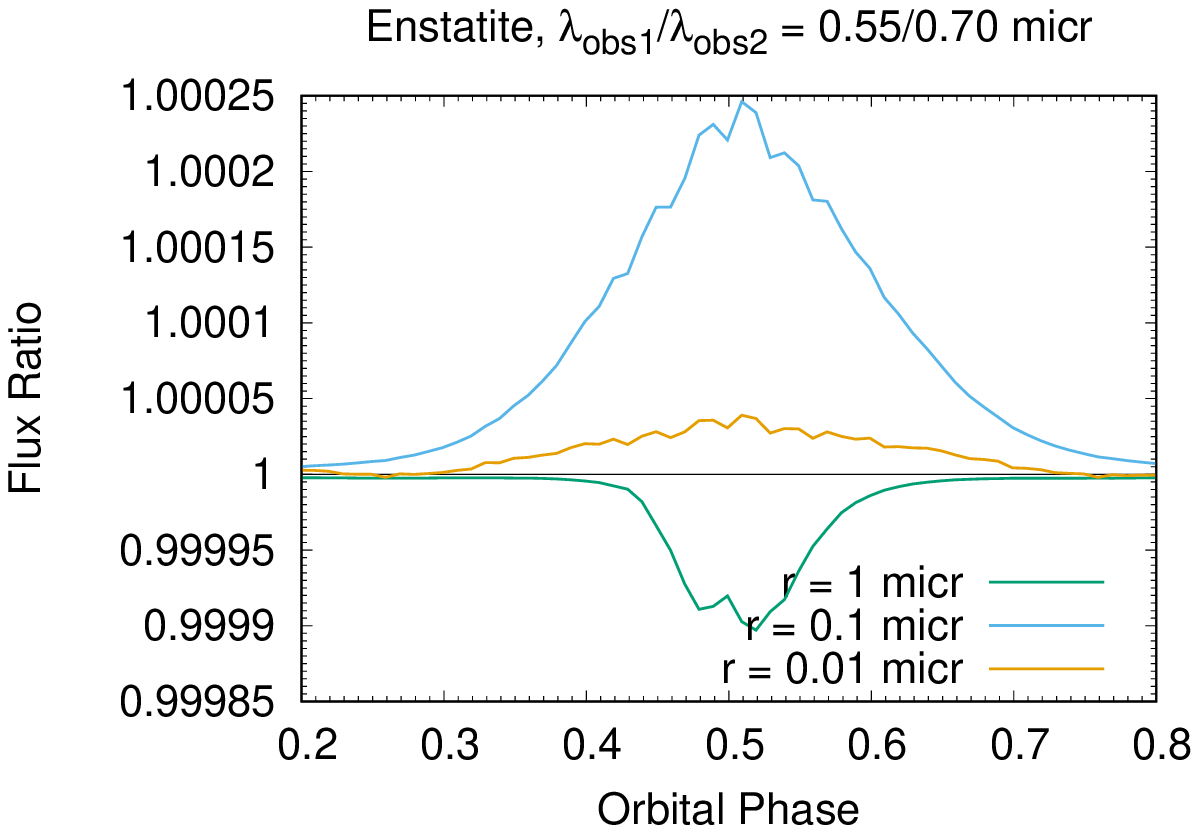}} 
\centerline{
\includegraphics[width=60mm]{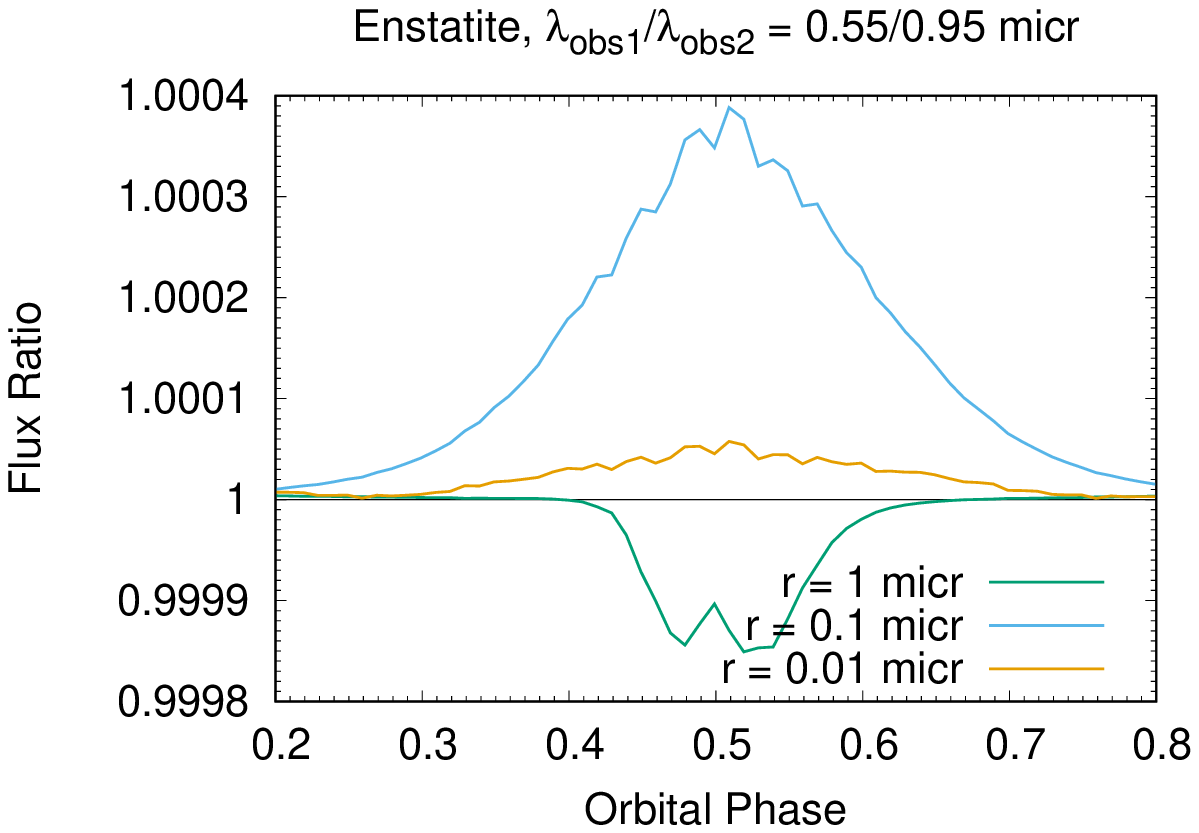} 
\includegraphics[width=60mm]{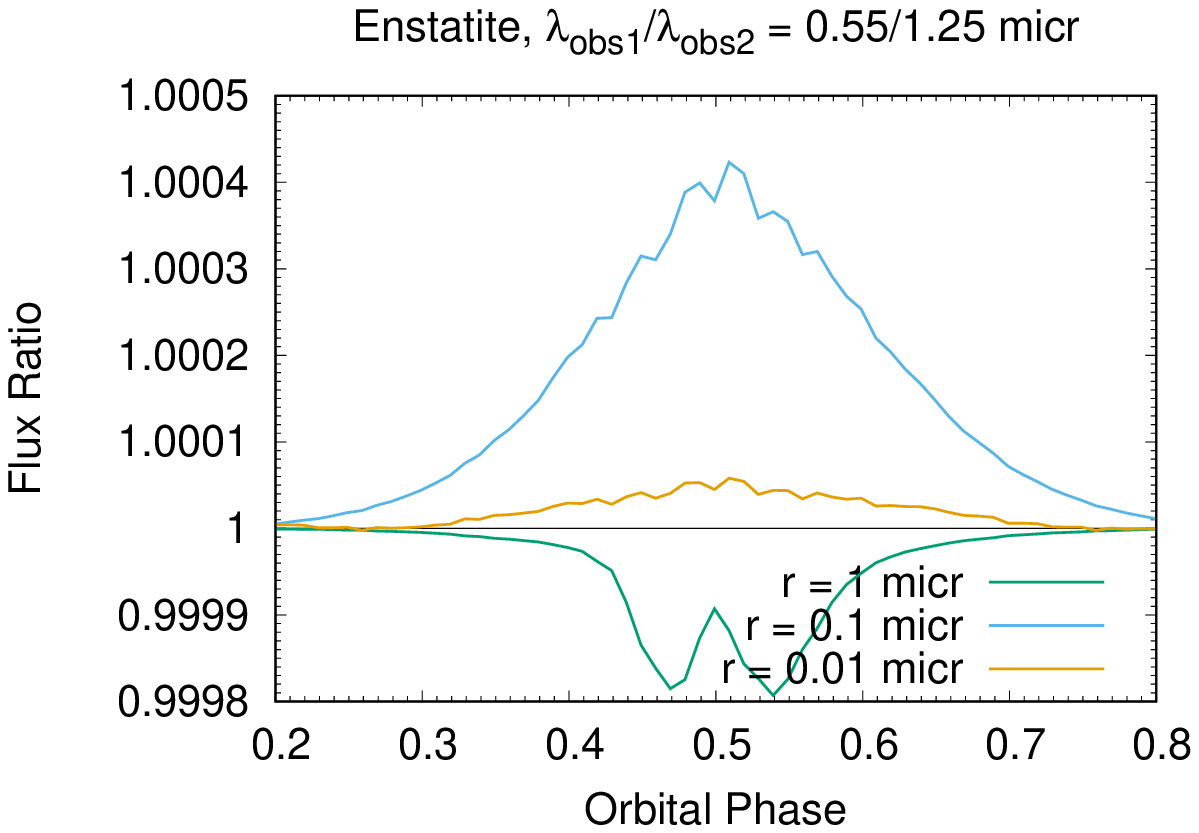}} 
\caption{Model light curves of the disintegrating exoplanet Kepler-1520b in a grazing, non-transiting regime with forward scattering only. Predicted photometry for different dust species and particle sizes, calculated for the \textit{Ariel} 0.55-micron wavelength observational channel. Flux ratios between \textit{Ariel} 0.55- and 0.70-micron, 0.55- and 0.95-micron, 0.55- and 1.25-micron wavelength observational channels are also shown. The same forward scattering can be applied in searching for grazing, non-transiting evaporating planets.}
\label{aluminafull}    
\end{figure*}

\begin{figure*}
\centering
\centerline{
\includegraphics[width=60mm]{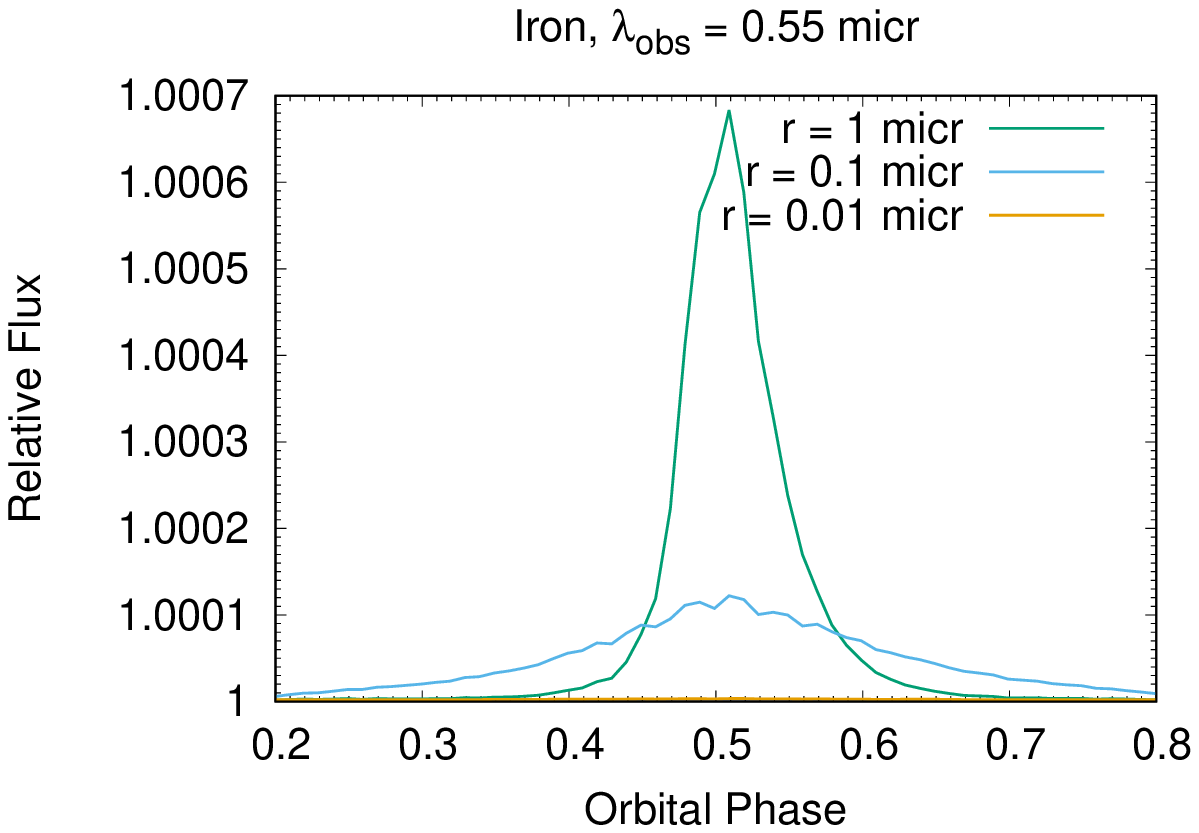} 
\includegraphics[width=60mm]{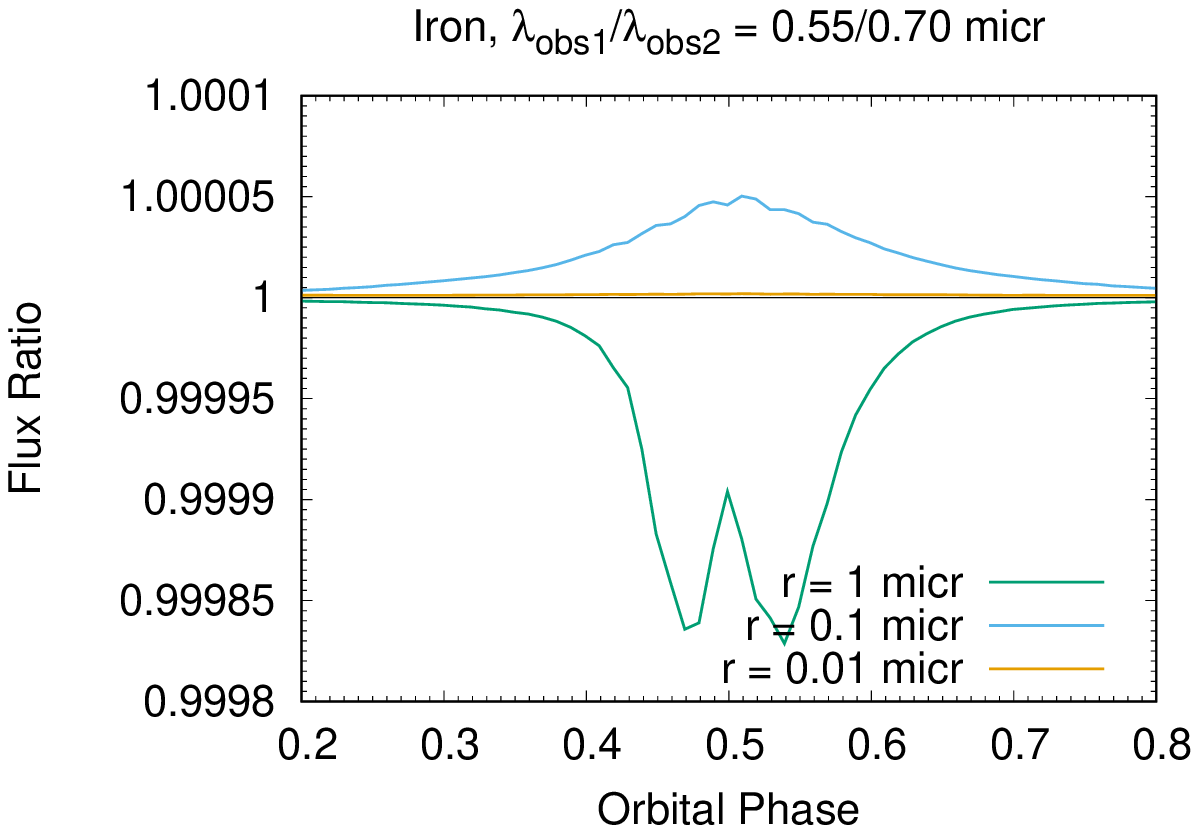}} 
\centerline{
\includegraphics[width=60mm]{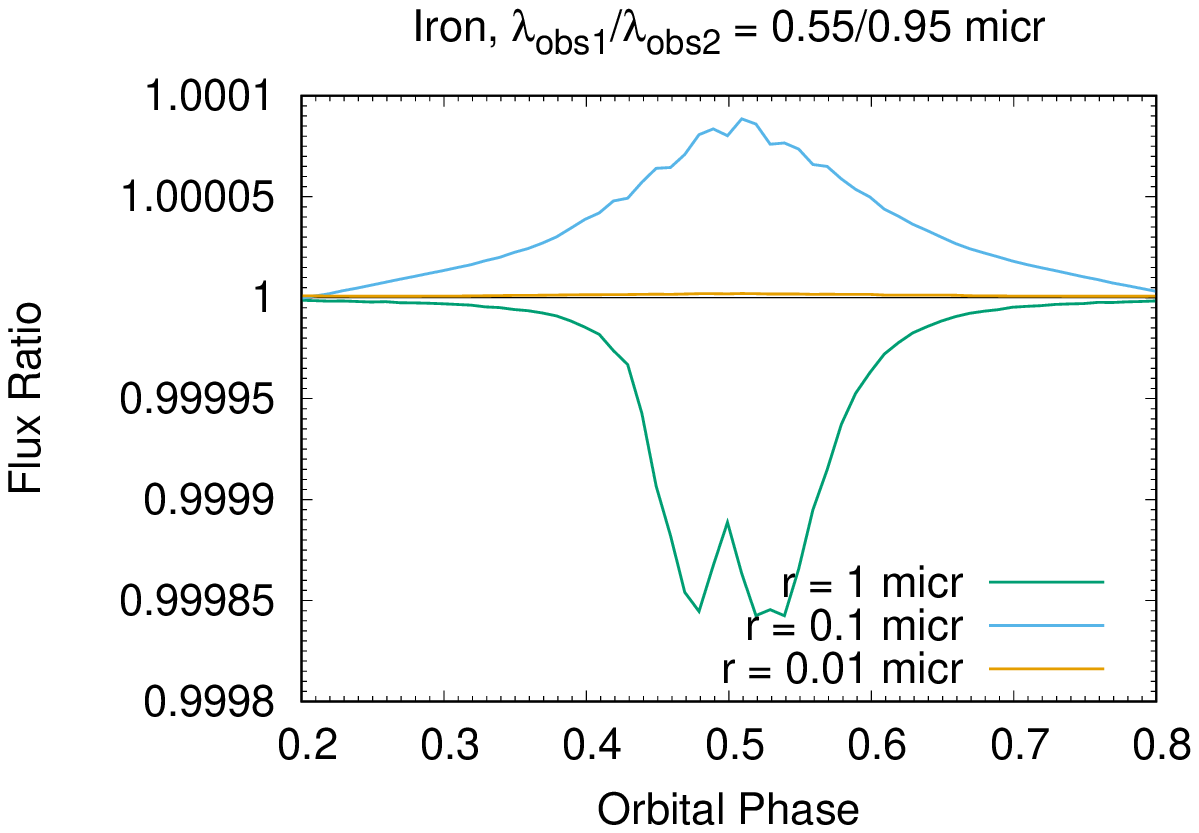} 
\includegraphics[width=60mm]{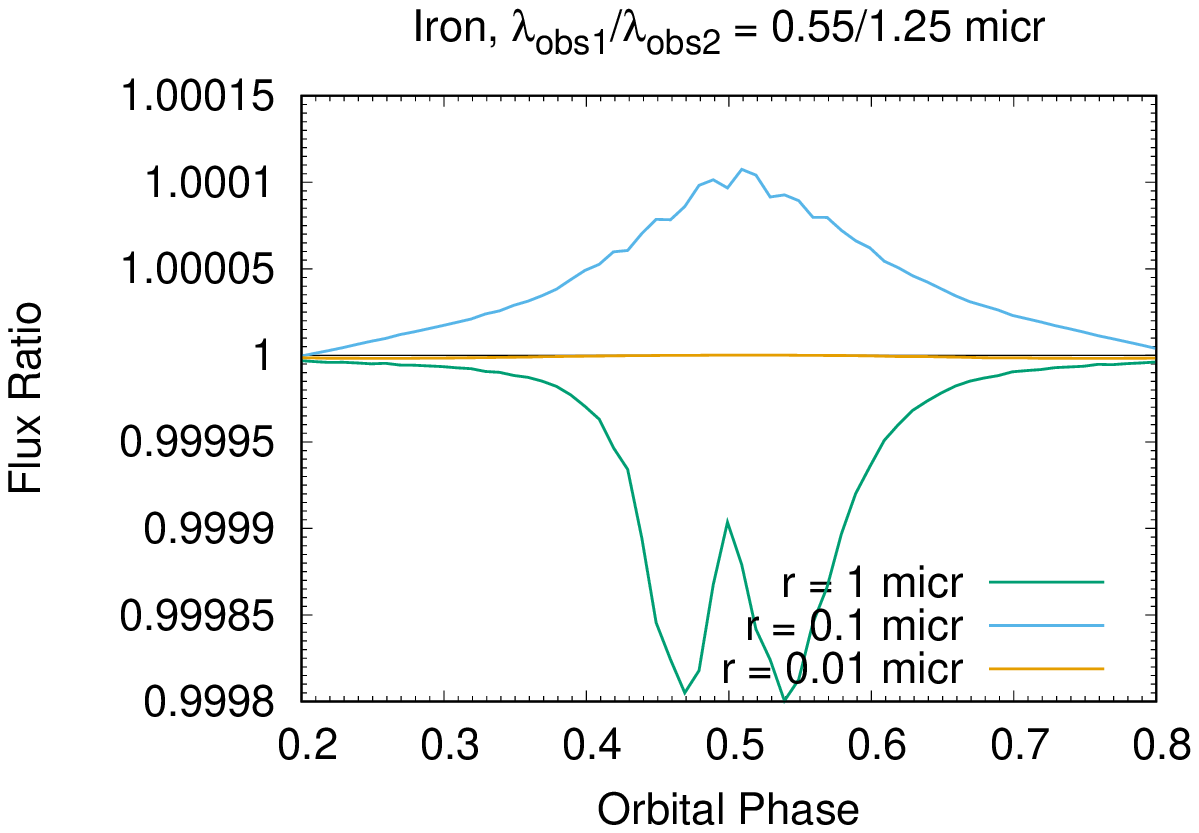}} 
\caption{As in Fig. \ref{aluminafull}, but for iron grains.}
\label{ironfull}    
\end{figure*}

To study these possibilities we took the orbital and planetary properties of Kepler-1520b discussed in Section \ref{transitmod}, and changed only the orbital properties of the system to get a grazing, non-transiting orbit scenario, where the solid body of the planet does not transit. Subsequently, we performed several modeling using the {\tt{Shellspec}} code. During this analysis we used the same optional object in the form of a modified ring as previously and assumed a spherical and limb-darkened central star with a radius of $R_\mathrm{s} = 0.66~\mathrm{R}_{\odot}$, mass of $M_\mathrm{s} = 0.76~\mathrm{M}_\odot$, and effective temperature of $T_\mathrm{eff} = 4677~\mathrm{K}$ \cite{Rappaport1}, located in the geometrical center of the ring. We modeled the comet-like tail as part of a ring with a radius of $a = 2.77~\mathrm{R}_{\odot}$. Its geometrical cross-section is monotonically enlarging from the planet to the end of the ring, which is located at $60^{\circ}$ behind the planet. We used the dust-tail cross-section of $0.05 \times 0.05~\mathrm{R}_\odot$ at the beginning and $0.09 \times 0.09~\mathrm{R}_\odot$ at its end. The density exponent $A2$ was fixed during the modeling procedure as previously, i.e., we used $A2 = -20$ in our calculations. The dust density at the beginning of the ring $\rho(0)$ was calculated previously for the given cross section of the tail, see Table \ref{rhovalues}. During this step we used these values also as fixed parameters. The key fixed parameter is, however, the orbit inclination angle $i$. By taking into account the above mentioned properties of the model we calculated that a non-transiting orbit scenario is possible if the orbit inclination angle is $i \leq 75^{\circ}$. In our analysis we finally decided to use the value of $i = 75^{\circ}$, which means that our hypothetical planet Kepler-1520b' is just non-transiting. We call this situation as "grazing, non-transiting scenario". We note that part of the dust-tail is still transiting, however, this is not its densest part, responsible for the observed forward scattering. The calculations were performed on the wavelengths of 0.55, 0.70, 0.95, 1.25, and 1.65 microns for consistency with the planned \textit{Ariel} observational channels. We executed 90 modeling calculations -- one for each combination of wavelengths, species, and dust particle sizes. Since in this case we used only fixed parameters, we did not need any goodness-of-fit parameter, nor uncertainty determination. The corresponding model light curves, calculated for the \textit{Ariel} 0.55-micron wavelength observational channel are depicted on the selected panels of Figs. \ref{aluminafull} and \ref{ironfull}. For better effect visibility, flux ratios between two \textit{Ariel} observational channels are also shown on the further panels of the same figure. We note that models composed from alumina are very similar to olivine- and pyroxene-models, enstatite- and forsterite-models are also very similar, and models calculated on 1.25 and 1.65 microns do not differ significantly, as well.                              

\begin{figure*}
\centering
\centerline{
\includegraphics[width=60mm]{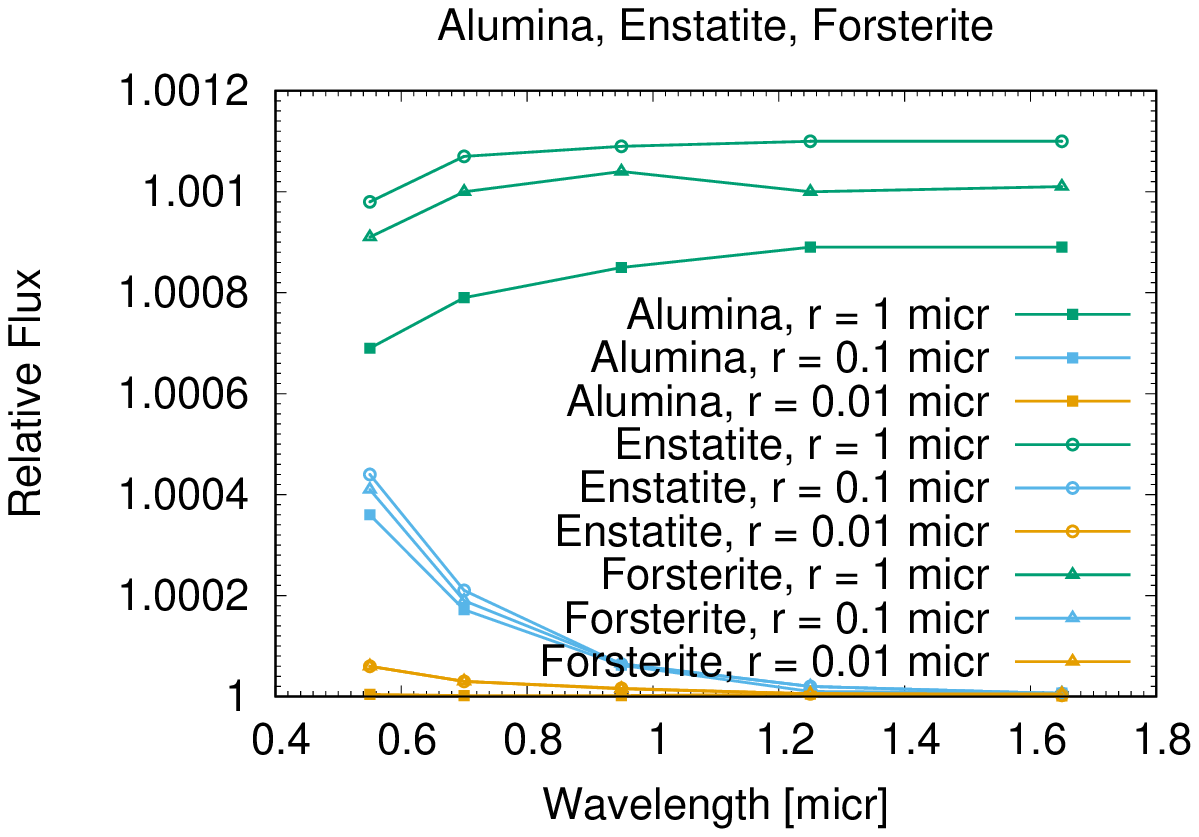} 
\includegraphics[width=60mm]{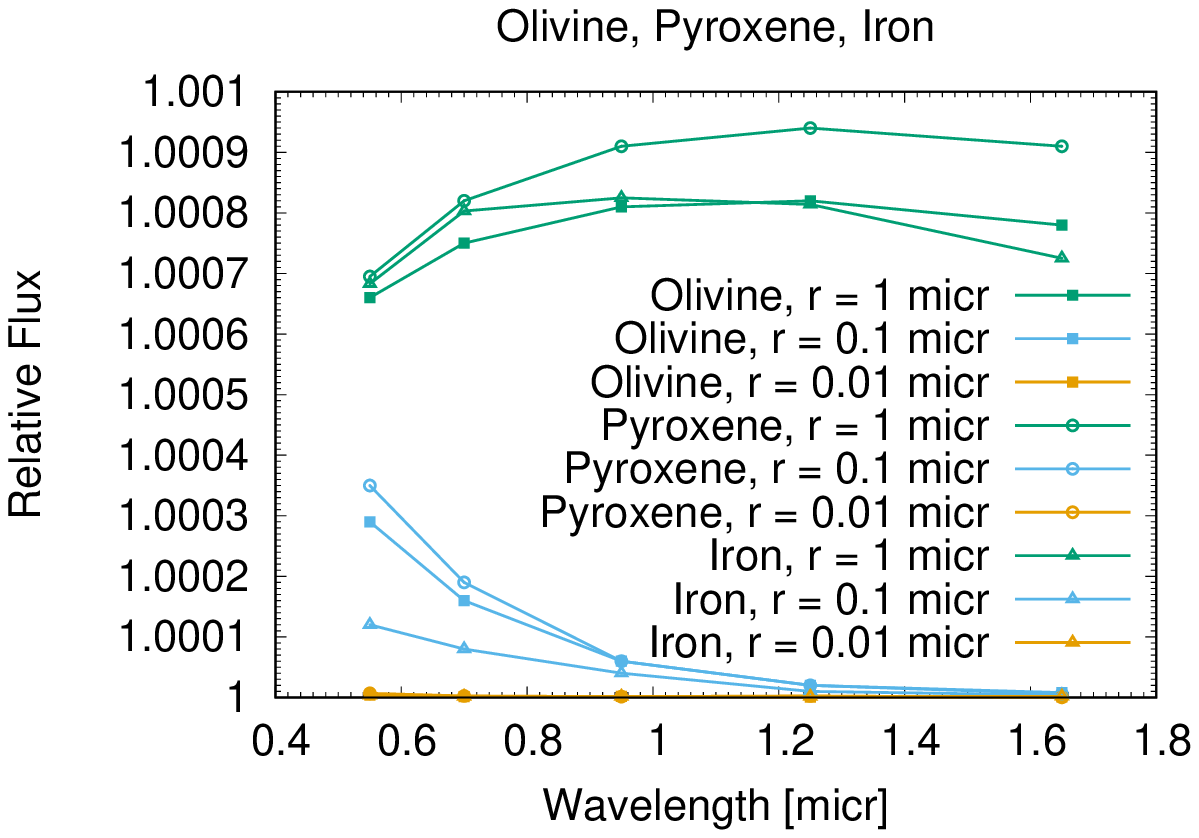}} 
\caption{Comparison of the forward-scattering amplitudes of alumina, enstatite, forsterite (left-hand panel), olivine, pyroxene, and iron grains (right-hand panel), calculated for the \textit{Ariel} observational channels.}
\label{amplitude}    
\end{figure*}

Based on the obtained models we can make several conclusions. (1) The first and the most important conclusion is that there are no significant differences among the selected species. (2) More significant differences are in the forward-scattering amplitude, as it is well visible on the panels of Fig. \ref{amplitude}. We can see that a change in the observational channel affects mainly the scattering amplitude of 0.1-micron grains. The 0.1-micron model becomes less dominant with the increasing wavelength. The most dominant is the 1-micron model, these particles generate forward-scattering peaks with the largest amplitude. The amplitude is affected by the observational channel only slightly, even though it is somewhat less dominant at the shortest wavelength. Similar properties have 0.01-micron grains, but with the difference that this is the least dominant model. These conclusions are in agreement with Fig. 13, presented in \cite{Croll1} (dimensionless grain size). (3) We can make the second group of conclusions concerning the duration of the forward scattering. From this viewpoint we can divide the models as follows. In the first group is the 1-micron model. This model is narrow and the duration of the scattering event is from 0.2 to 0.4 in units of phase. In the second group are the 0.1- and 0.01-micron models. These models are not so narrow as the 1-micron model and they have wider wings near the continuum. The duration of the scattering event is from 0.4 to 0.6 in units of phase. 

About the detectability of a grazing, non-transiting disintegrating planet, creating a comet-like tail from the evaporating material we can conclude the followings. (1) The most difficult is to detect evaporating material composed from 0.01-micron grains, which has the smallest forward-scattering peak at any of the \textit{Ariel} wavelengths. (2) If the evaporating material is composed from 0.1-micron grains, observations on 0.55 micron is the best way to detect such a disintegrating planet. (3) In the most ideal case the evaporating material is composed from 1-micron grains, which could be the most detectable material at any of the \textit{Ariel} wavelengths. For example, \cite{Brogi1} found that the typical particle size in the comet-like tail of Kepler-1520b (near the planet) is 0.1 micron. If Kepler-1520b would be a grazing, non-transiting planet with the same planet properties we could detect this object using \textit{Ariel} very probably only at 0.55 micron. On the other hand, it is clear that we do not have information about the grain size of unknown disintegrating planets, therefore multiwavelength observations are needed to increase the probability of detections.             

\section{Discussion}
\label{discuss}

In the previous section we did not take into account several accompanying factors, which could affect the detectability of a grazing, disintegrating planet. In this section we discuss three of them, which we consider as very important: white noise, the orbit inclination angle, and the broad-band nature of the photometry.  

\subsection{White noise}

\begin{figure*}
\centering
\centerline{
\includegraphics[width=60mm]{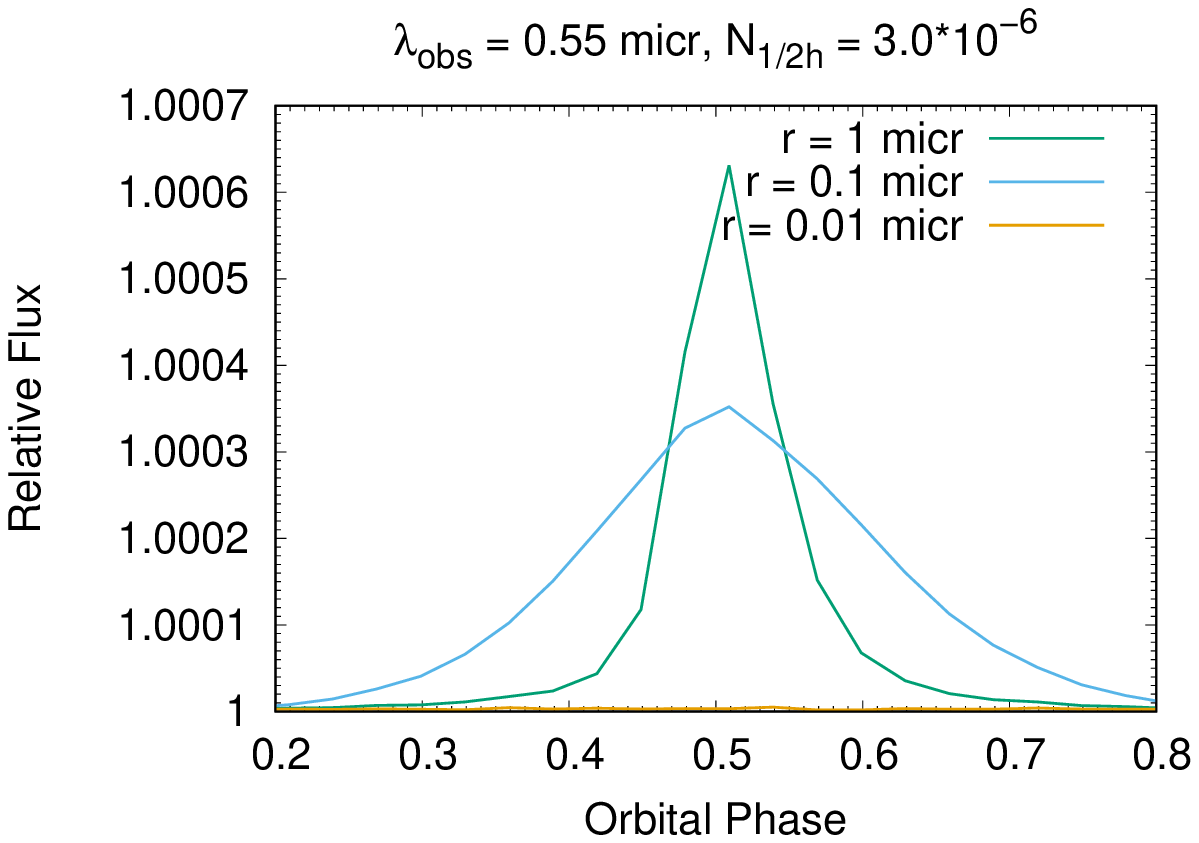} 
\includegraphics[width=60mm]{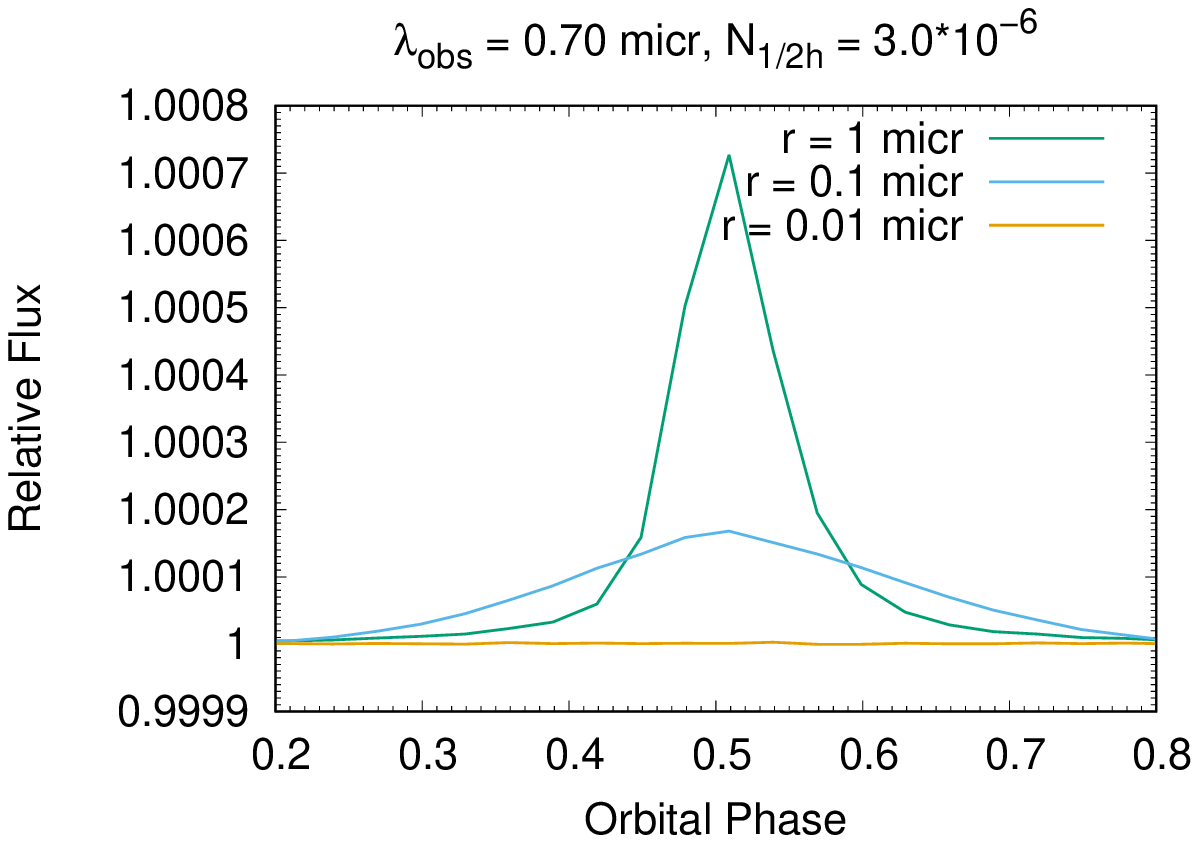}} 
\centerline{
\includegraphics[width=60mm]{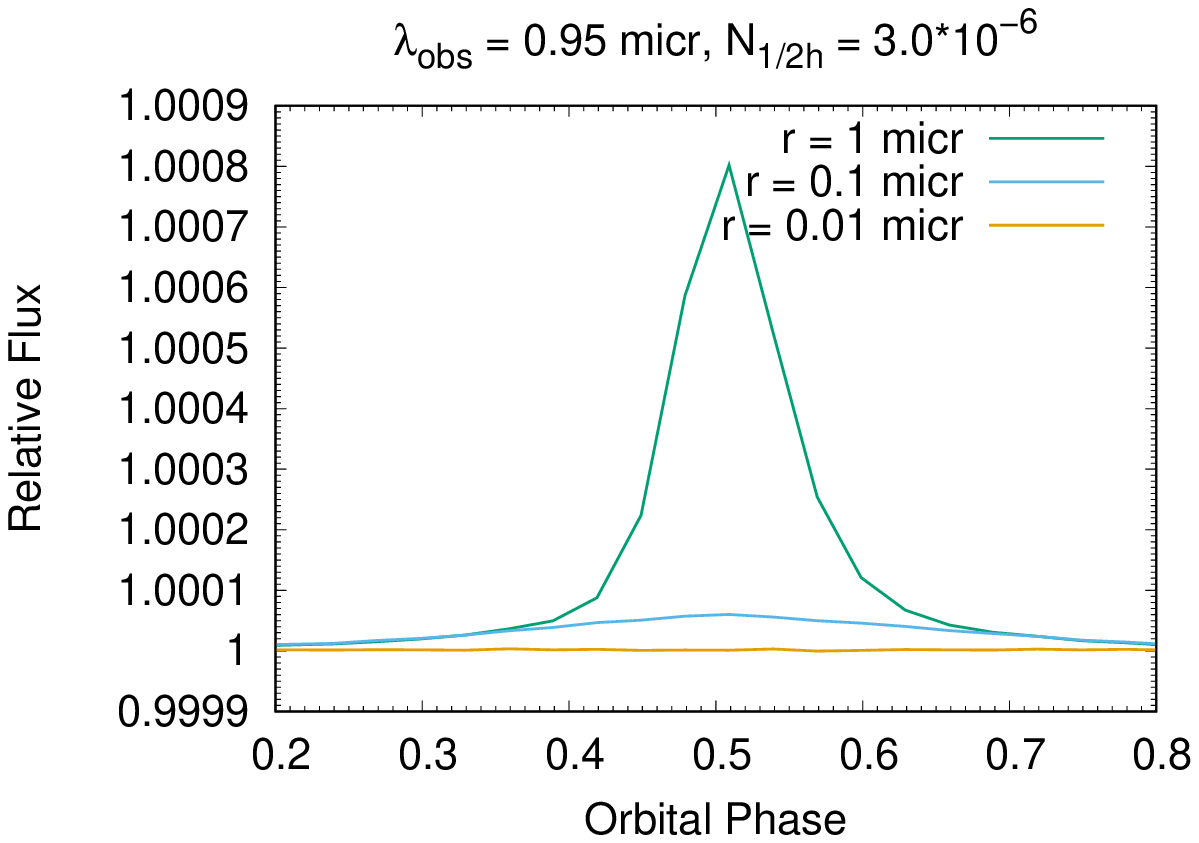} 
\includegraphics[width=60mm]{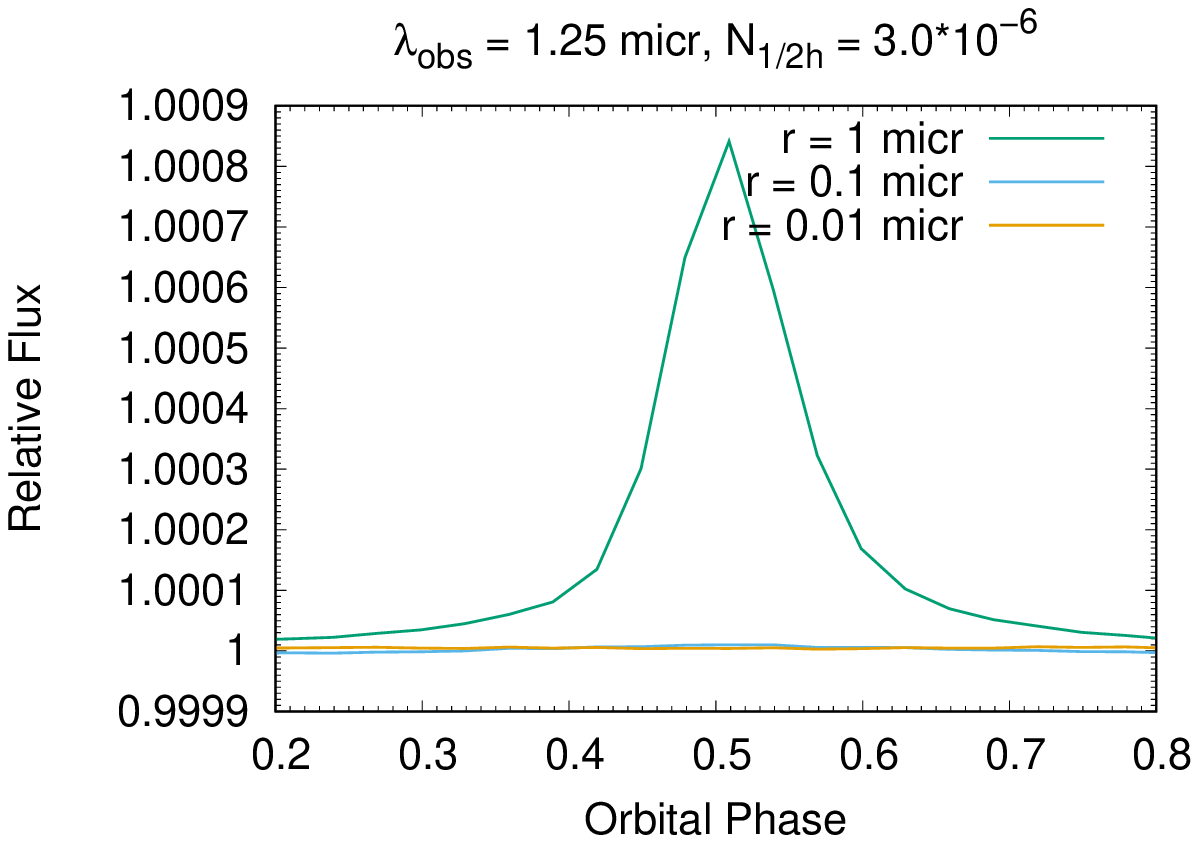}} 
\centerline{
\includegraphics[width=60mm]{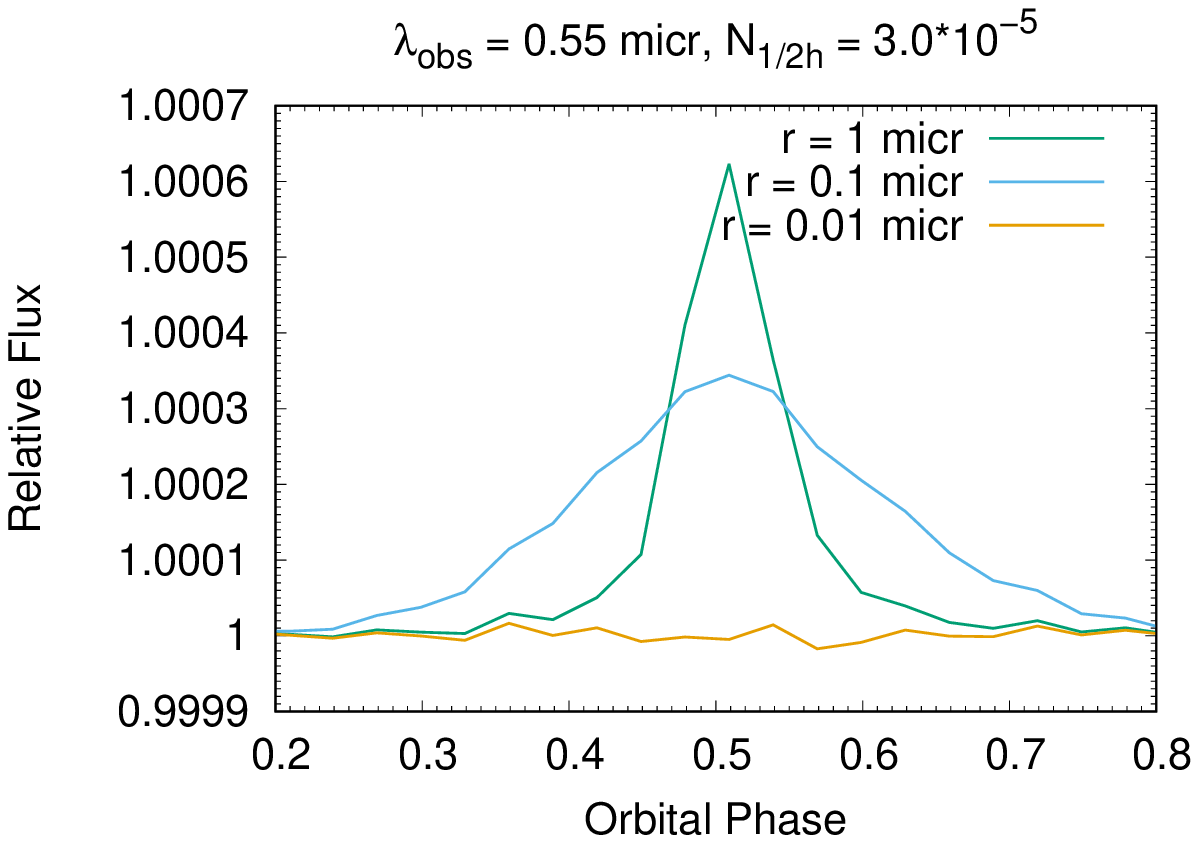} 
\includegraphics[width=60mm]{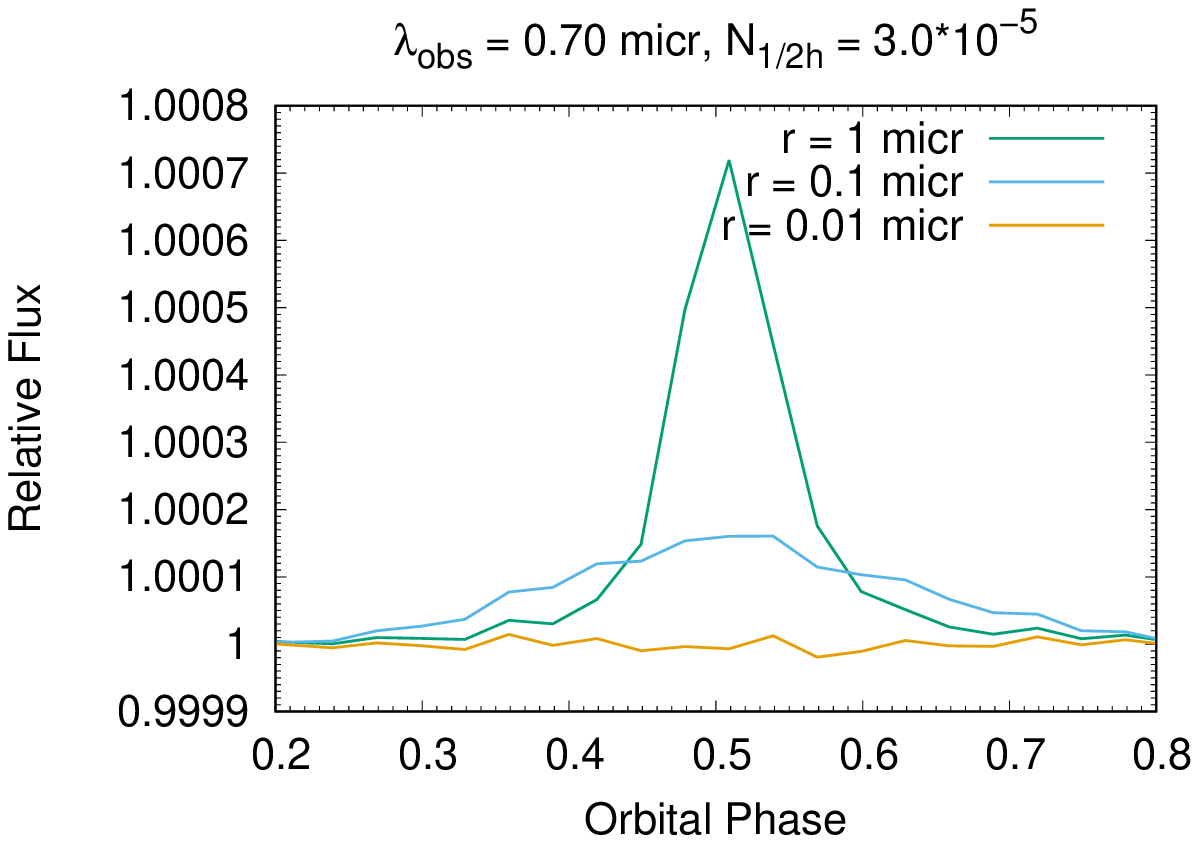}} 
\centerline{
\includegraphics[width=60mm]{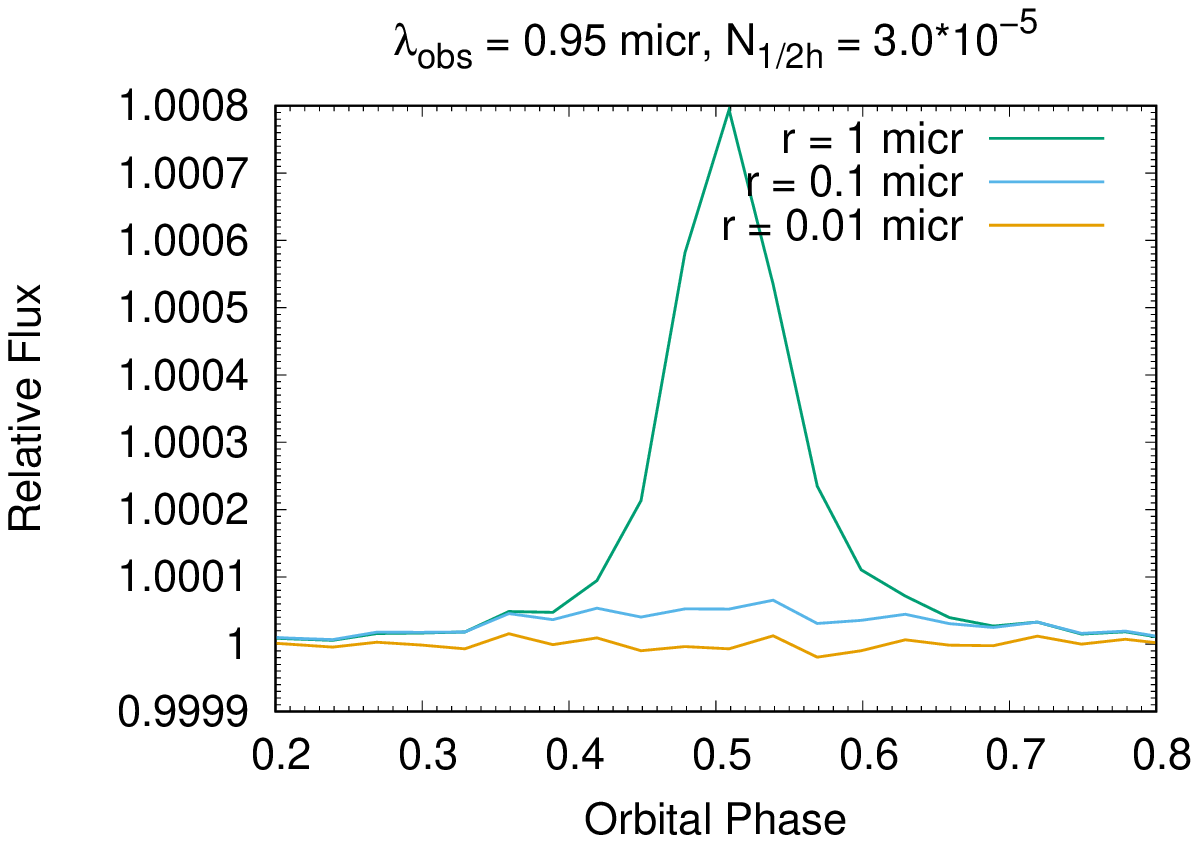} 
\includegraphics[width=60mm]{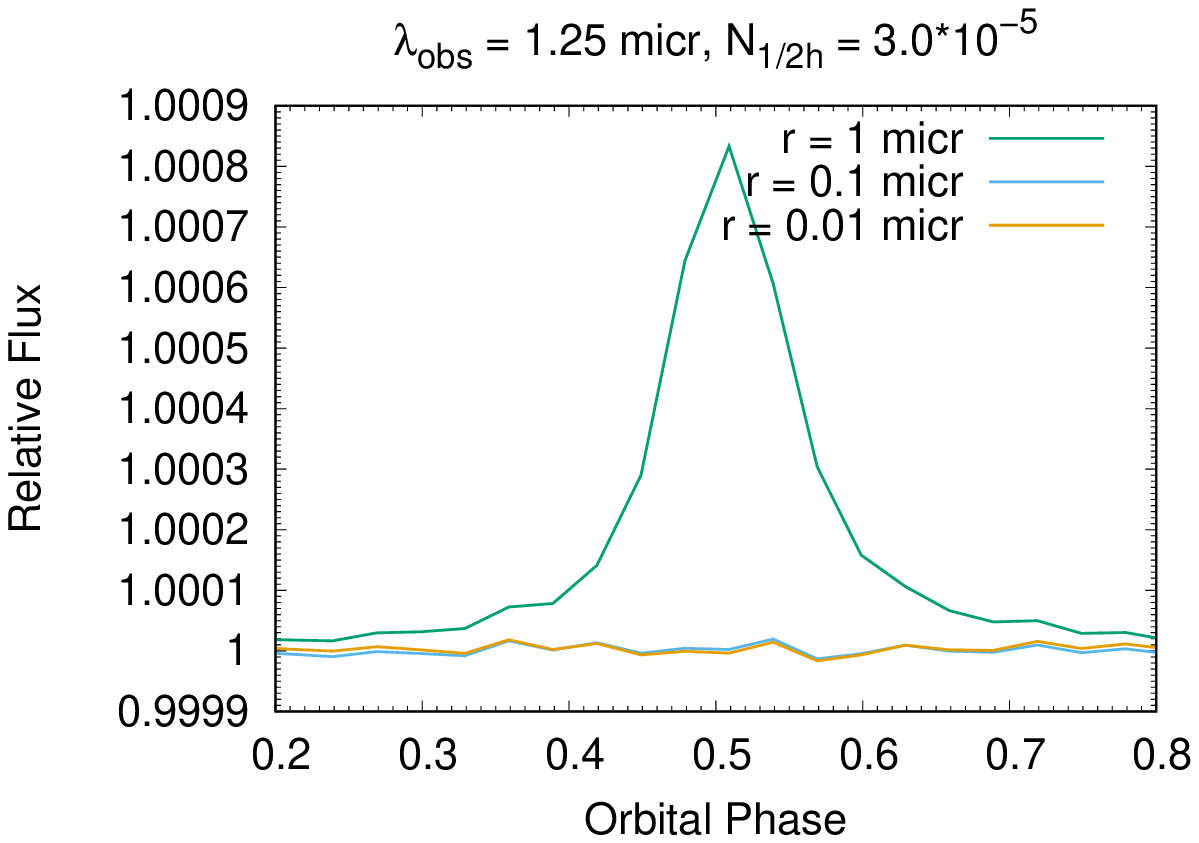}} 
\caption{Model light curves of the disintegrating exoplanet Kepler-1520b in a grazing, non-transiting regime with forward scattering and white noise (part I). Predicted photometry for different \textit{Ariel} observational channels and particle sizes of alumina.}
\label{noise106}    
\end{figure*}

\begin{figure*}
\centering
\centerline{
\includegraphics[width=60mm]{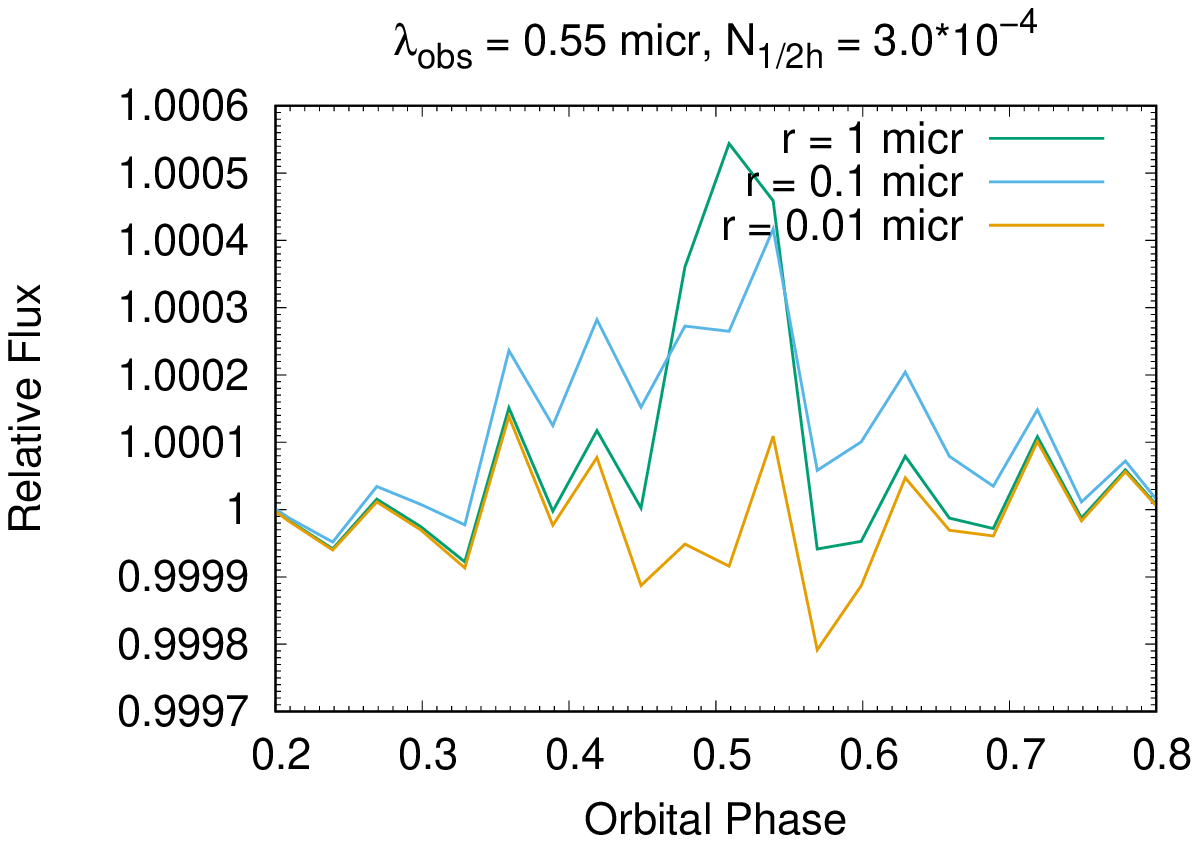} 
\includegraphics[width=60mm]{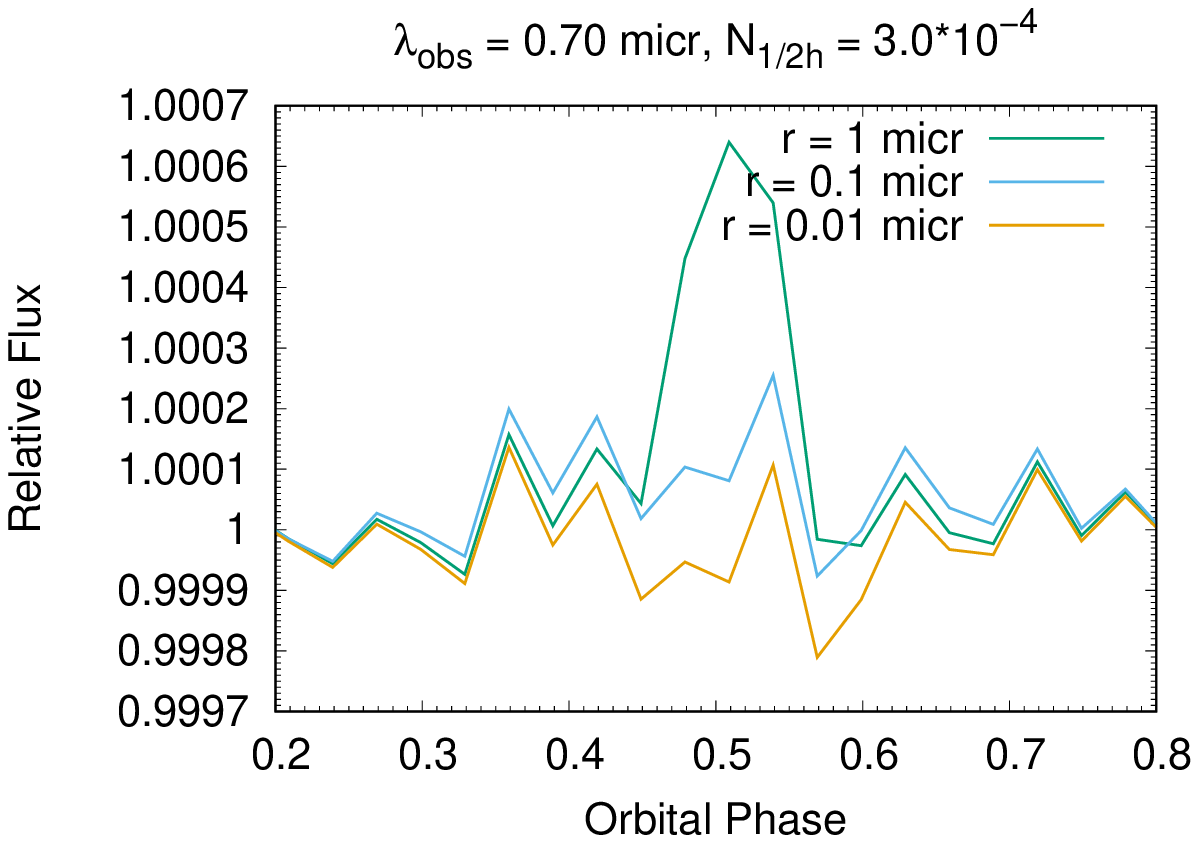}} 
\centerline{
\includegraphics[width=60mm]{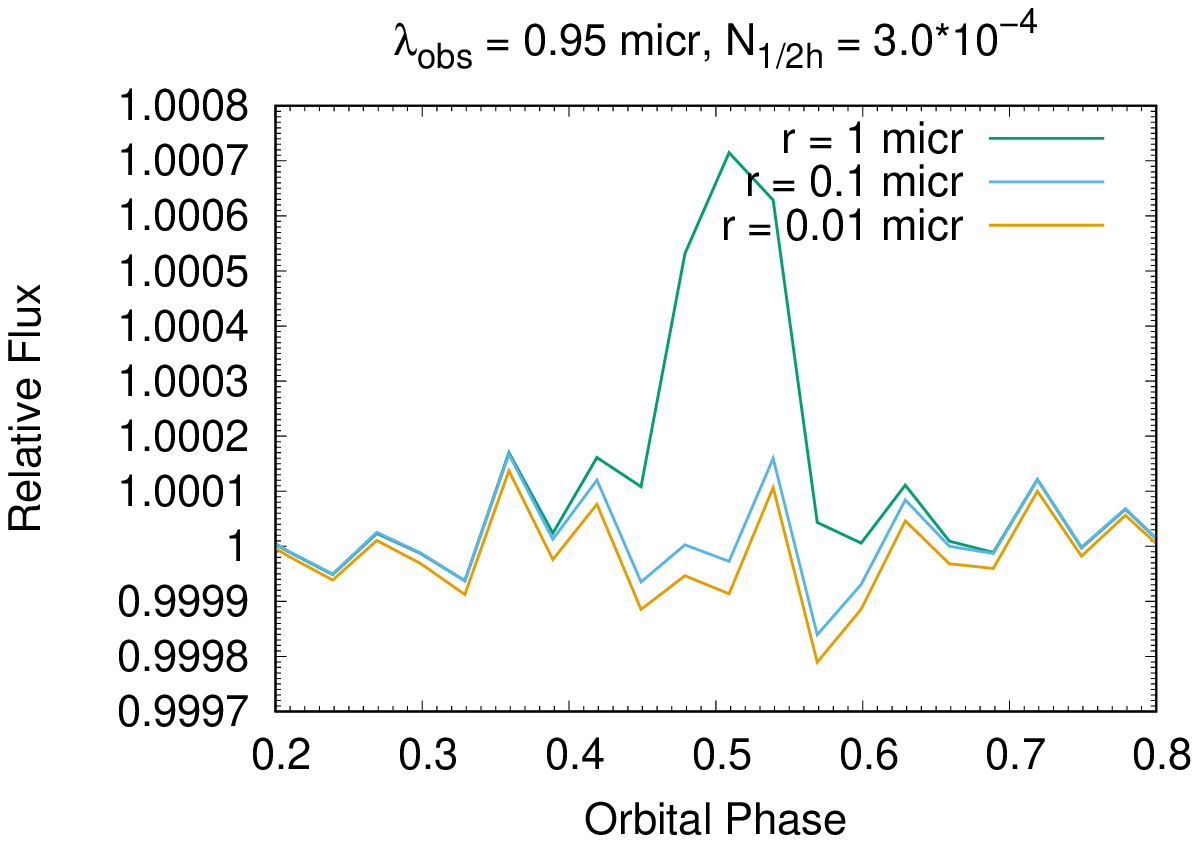} 
\includegraphics[width=60mm]{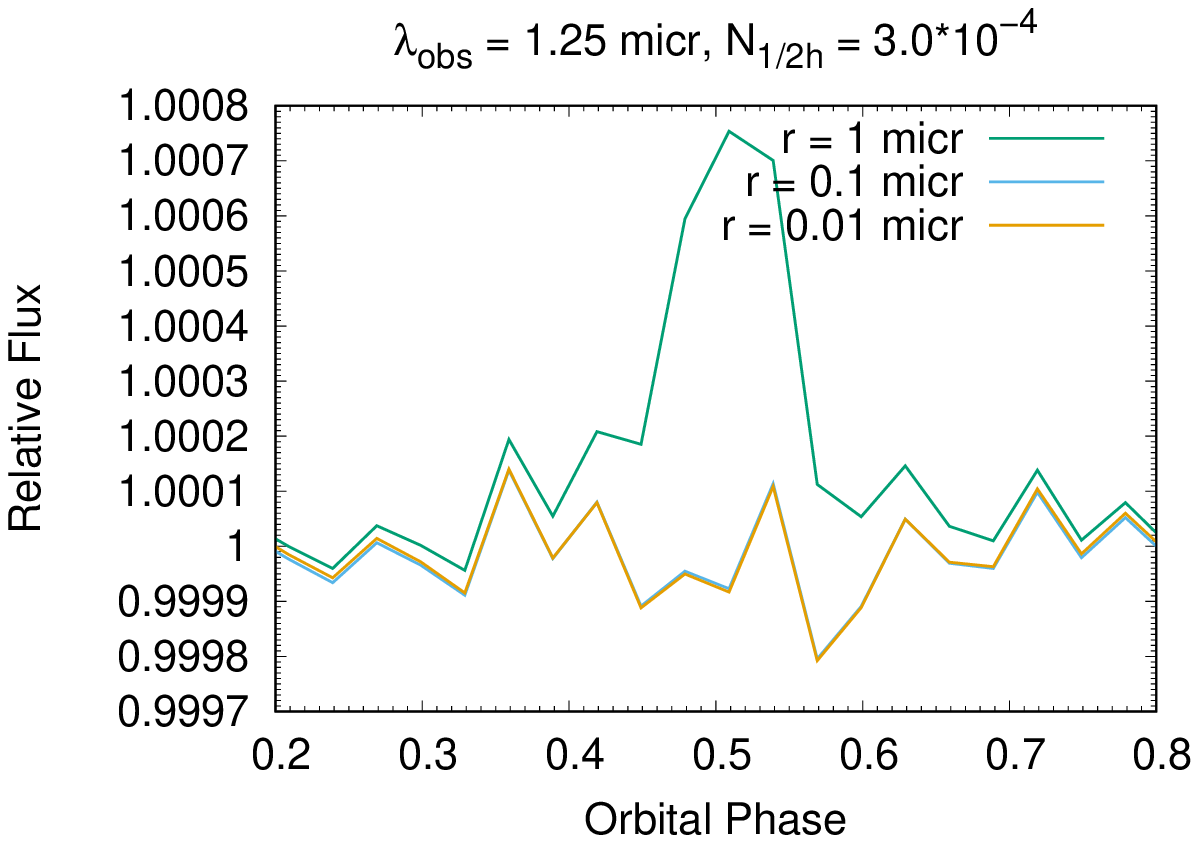}} 
\caption{Model light curves of the disintegrating exoplanet Kepler-1520b in a grazing, non-transiting regime with forward scattering and white noise (part II). Predicted photometry for different \textit{Ariel} observational channels and particle sizes of alumina.}
\label{noise104}    
\end{figure*}

It is very important limiting factor at photometric observations. Since no significant differences among the selected species, we worked only with the models calculated for alumina. We added white noise in these models. White noise has zero mean, constant variance, and is uncorrelated in time. Since the tabulated \textit{Ariel} noise parametrisation\footnote{Obtained from Gy. M. Szab\'{o} via private communication, \url{szgy@gothard.hu}.} gives the total white noise calculated for 30-min integration, we rebinned the model data using the 30-min-width window, and tried the following 30-min noise levels: $N_\mathrm{1/2h} = 3 \times 10^{-6}$ (3 ppm), $3 \times 10^{-5}$ (30 ppm), $3 \times 10^{-4}$ (300 ppm), $3 \times 10^{-3}$ (3000 ppm), $3 \times 10^{-2}$ (30~000 ppm = 3\%), and $3 \times 10^{-1}$ (300~000 ppm = 30\%). The corresponding model light curves are depicted on the panels of Figs. \ref{noise106} and \ref{noise104}. We note that the models calculated on 1.25 and 1.65 microns do not differ significantly, also the models with white noise of $N_\mathrm{1/2h} \geq 3 \times 10^{-3}$ ($\geq 3000$ ppm) are below the detection limit. Based on these models we can conclude the followings. (1) If the evaporating material is composed from 1-micron grains of alumina, the 30-min integrated noise in the data should be $N_\mathrm{1/2h} \leq 10^{-4}$ ($\leq 100-900$ ppm) to detect a Kepler-1520b-like disintegrating planet in a grazing, non-transiting regime. In this case the observational wavelength does not matter. (2) If the evaporating material is composed from 0.1-micron grains of alumina, the 30-min integrated noise in the data should be $N_\mathrm{1/2h} \leq 10^{-5}$ ($\leq 10-90$ ppm) for a convincing detection. In this case the detectability is strongly affected by the observational wavelength. As we concluded in Section \ref{grazingmod}, observations on 0.55 micron could be the most productive. (3) Evaporating material composed from 0.01-micron grains of alumina is not possible to detect nor with noise of $N_\mathrm{1/2h} = 3 \times 10^{-6}$ (3 ppm). 

We can note that other species could have similar properties of detectability. In the case of Kepler-1520b in a grazing, non-transiting scenario with the typical particle size of 0.1 micron, observed on 0.55 micron, the 30-min integrated noise in the data should be $N_\mathrm{1/2h} \leq 10^{-5}$ ($\leq 10-90$ ppm) for a convincing detection. We note, however, that the calculations do not contain red noise, which has zero mean, constant variance, and is serially correlated in time, therefore the real precision needed for detections may be higher. On the other hand, longer photometric runs registering several orbits of such a disintegrating exoplanet could improve the observational precision via phase-folding the obtained data. Based on the tabulated \textit{Ariel} noise parametrisation\footnote{The expected noise spans from $N_\mathrm{1/2h} = 9 \times 10^{-4}$ (900 ppm) at the faintest stars ($K = 13.5$ mag), up to $N_\mathrm{1/2h} = 4 \times 10^{-5}$ (40 ppm) at the brightest stars ($K = 3.6$ mag).} we can conclude that the predicted white noise, calculated for 30-min integration seems to be enough to detect some disintegrating exoplanets in a grazing, non-transiting regime, as well. \textit{Ariel} core survey will monitor about 1000 stars and the list of potential targets can be found in \cite{Edwards1}. If we consider only $N_\mathrm{1/2h} \leq 10^{-5}$ ($\leq 10-90$ ppm, $K < 10.0$ mag, $d < 100$ pc) for a convincing detection, roughly 50\% of the \textit{Ariel} core survey targets\footnote{See top left-hand panel of Fig. 2 in \cite{Edwards1}.} might be expected to have sufficient signal-to-noise ratio for detection of a Kepler-1520b-like disintegrating planet in a grazing, non-transiting regime. This optimistic number will be, however, reduced by other factors, as well (e.g., by the orbit inclination angle).

\subsection{The orbit inclination angle}

\begin{figure*}
\centering
\centerline{
\includegraphics[width=60mm]{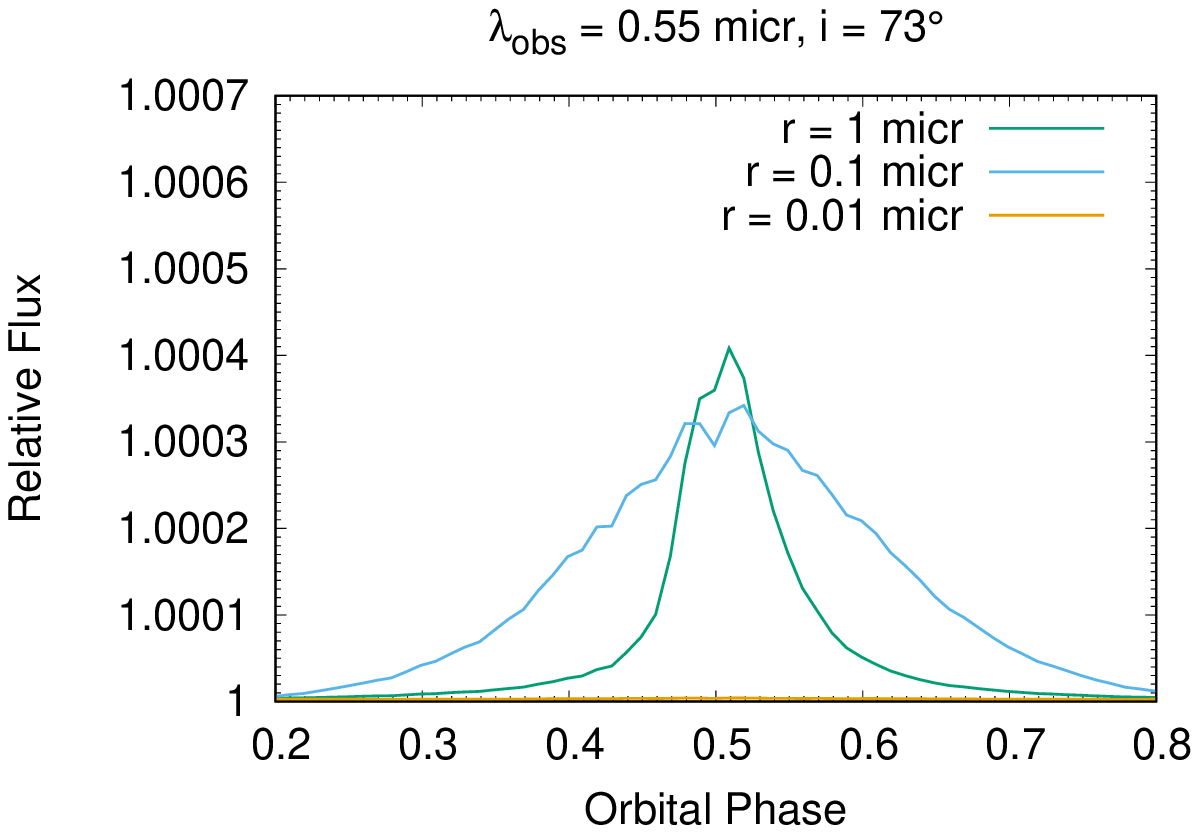} 
\includegraphics[width=60mm]{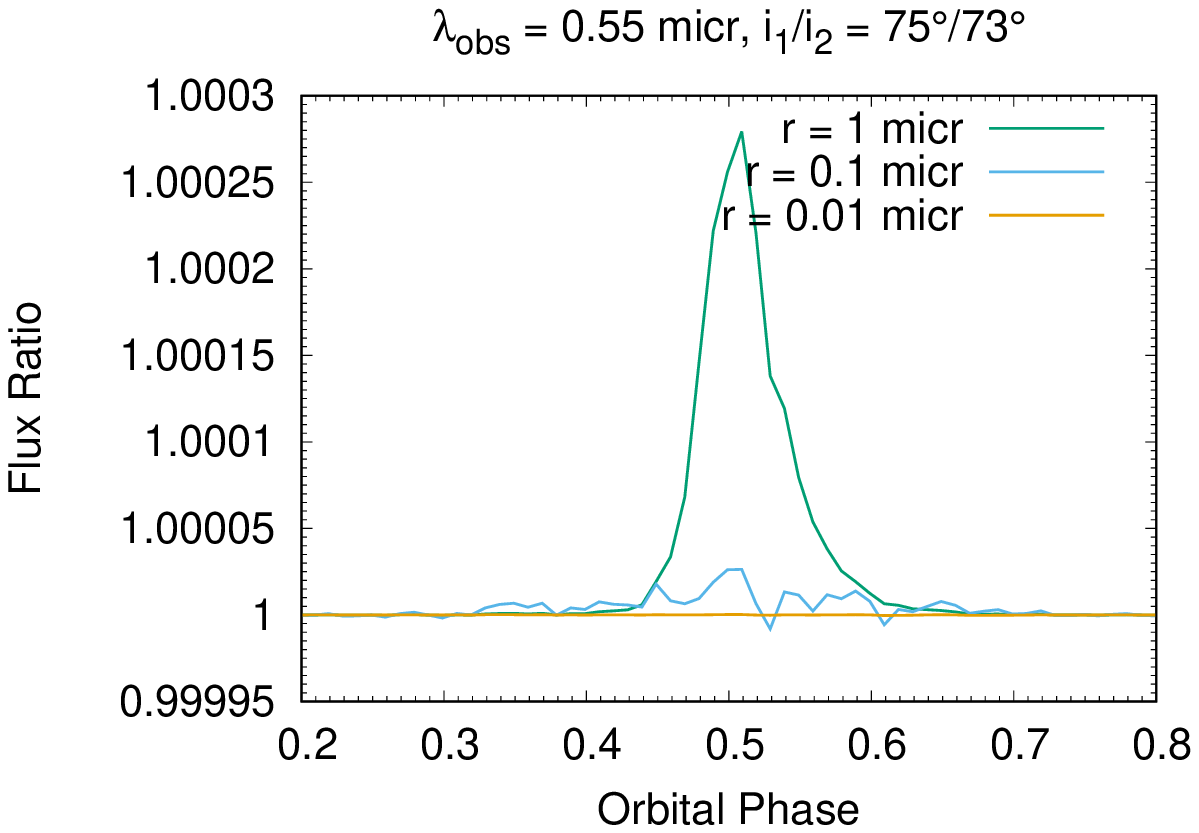}} 
\caption{Model light curves of the disintegrating exoplanet Kepler-1520b in a grazing, non-transiting regime with forward scattering only. Predicted photometry for different particle sizes of alumina, calculated for the orbit inclination angle of $i = 73^{\circ}$, and for the \textit{Ariel} wavelength of 0.55 micron (left-hand panel). Flux ratios between the light curves calculated for $i = 75^{\circ}$ and $i = 73^{\circ}$ are also shown (right-hand panel).}
\label{incl}    
\end{figure*}

Planetary systems also have one more degree of freedom, i.e., the orbit inclination angle. In the previous section we investigated only the case of $i = 75^{\circ}$, but as \cite{DeVore1} pointed out, the amplitude of the forward scattering decreases with decreasing inclination angle. To illustrate the effect of a change in the orbit inclination angle value on the light curves we selected the models, which were calculated for alumina on 0.55 micron without white noise (see Fig. \ref{aluminafull}, top left-hand panel), changed only the orbit inclination angle value from $i = 75^{\circ}$ to $i = 73^{\circ}$, and recalculate these models. The recalculated model light curves are depicted on the left-hand panel of Fig. \ref{incl}. For better effect visibility, flux ratios between the original and the recalculated light curves are also shown on the right-hand panel of the same figure. 

\begin{figure*}
\centerline{
\includegraphics[width=60mm]{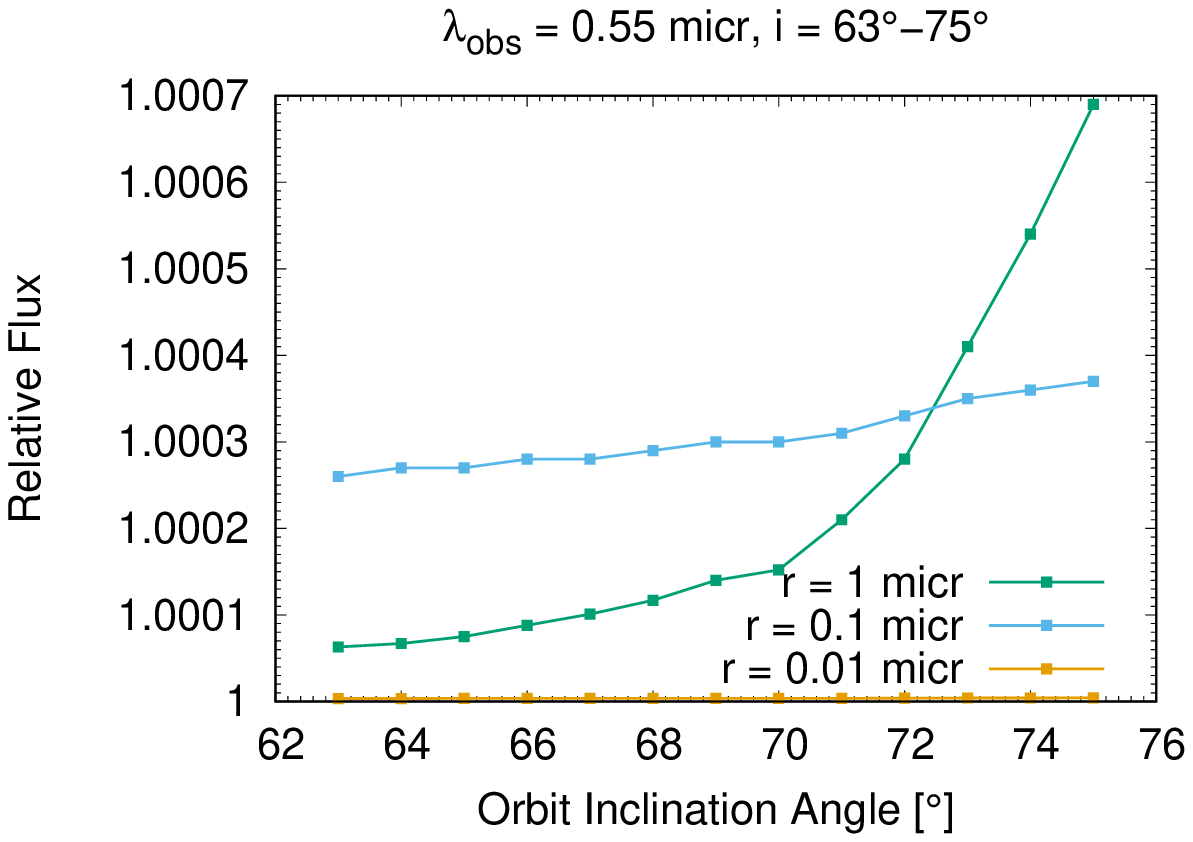} 
\includegraphics[width=60mm]{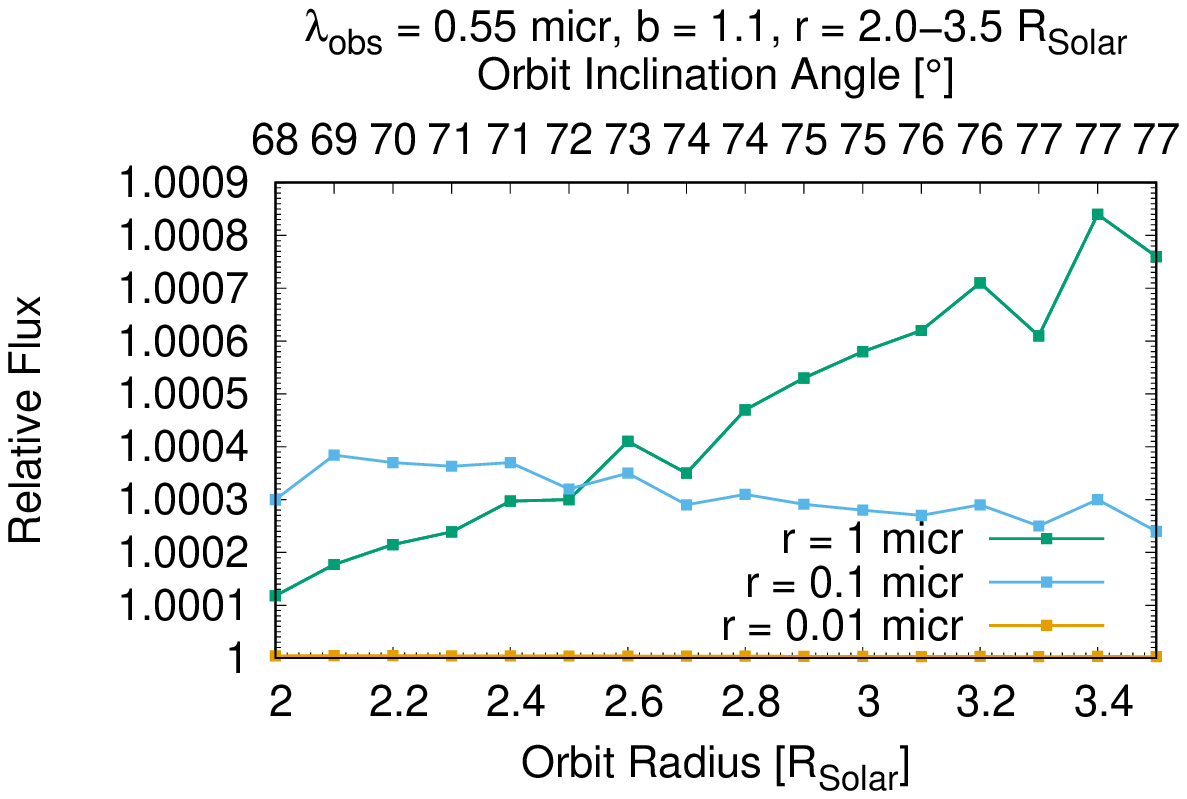}} 
\caption{Comparison of the forward-scattering amplitudes of alumina grains, calculated for the \textit{Ariel} wavelength of 0.55 micron, and for different orbit inclination angles (left-hand panel). The same comparison, but the orbit radius is also changing, at the same time the impact parameter remains roughly constant (right-hand panel).}
\label{inclandimpact}    
\end{figure*}

Although this is only an illustrative example, we can clearly see, how this small change of $\Delta i = 2^{\circ}$ in the orbit inclination angle affects the light curves. This factor changed mainly the scattering amplitude of the 1-micron model. The 0.1-micron model calculated at $i = 73^{\circ}$ is very similar to the original model with the orbit inclination angle of $i = 75^{\circ}$. This is also true for the 0.01-micron model, even though this light curve is below the detection limit. We can also see that now the models calculated for 0.1- and 1-micron grains have comparable amplitude. We can distinguish them based on the "sharpness" of the light curves. In Section \ref{grazingmod} we concluded that in the most ideal case the evaporating material is composed from 1-micron grains, which could be the most detectable material at any of the \textit{Ariel} wavelengths. Here we can see (Fig. \ref{inclandimpact}, left-hand panel) that this is valid only for the case, if the disintegrating planet is just non-transiting. With decreasing inclination angle the forward-scattering amplitude of 1-micron grains decreases rapidly, as well. The forward-scattering amplitude of 0.1-micron grains has similar trend with decreasing inclination angle, but the change rate in the flux is smaller.

Assuming a circular orbit, if we change not only the orbit inclination angle, but also the orbit radius, at the same time the impact parameter, defined as the projected distance between the planet and the star centre, remains roughly constant, we can get the following results. As it is depicted on the right-hand panel of Fig. \ref{inclandimpact}, a change in the orbit radius/inclination angle affects mainly the model composed from 1-micron grains, which is less dominant at smaller distances from the parent star, but the forward-scattering amplitude increases with the orbit radius. A small decrease in the forward-scattering amplitude of 0.1-micron grains is visible with increasing orbit radius/inclination angle. The scattering amplitude of 0.01-micron grains is not affected by the orbit radius/inclination angle. We can therefore conclude that from this point of view grazing, non-transiting disintegrating planets with longer orbital periods could be favored over ultra-short planets.     

\subsection{The broad-band nature of the photometry}

\begin{figure*}
\centering
\centerline{
\includegraphics[width=60mm]{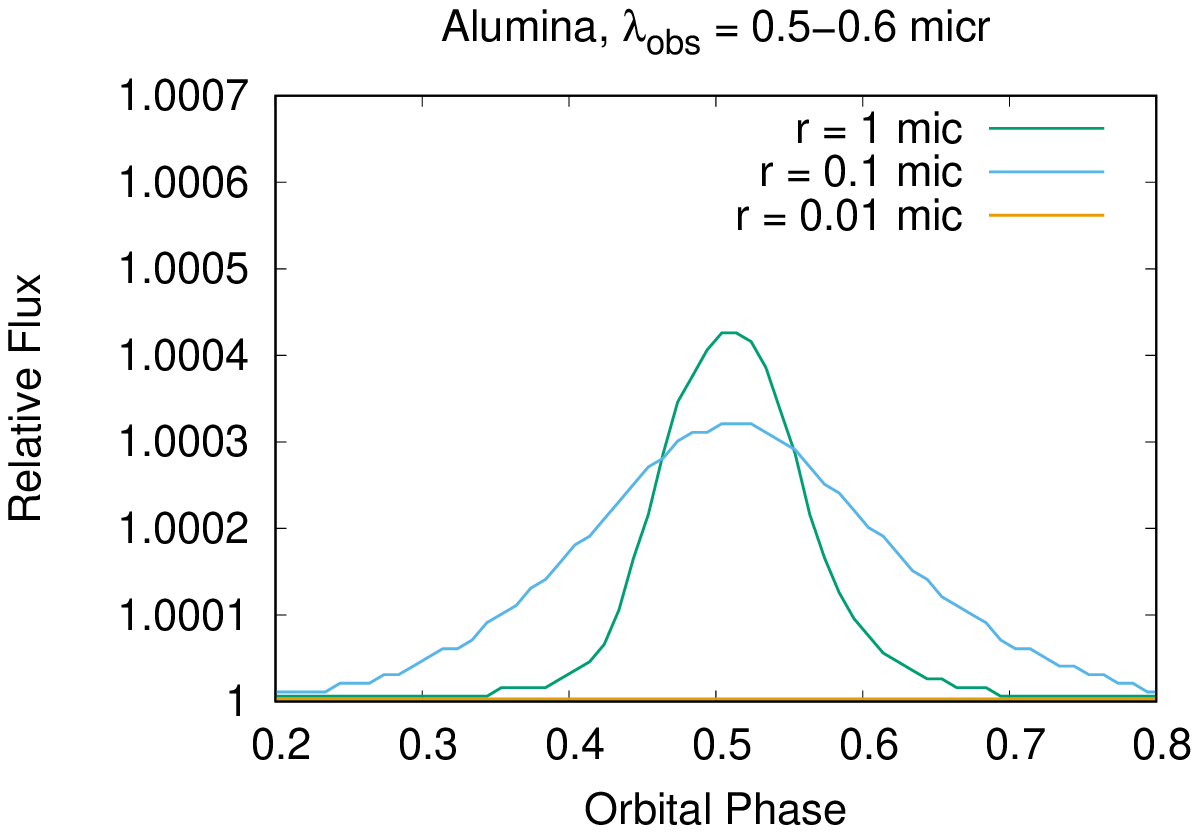} 
\includegraphics[width=60mm]{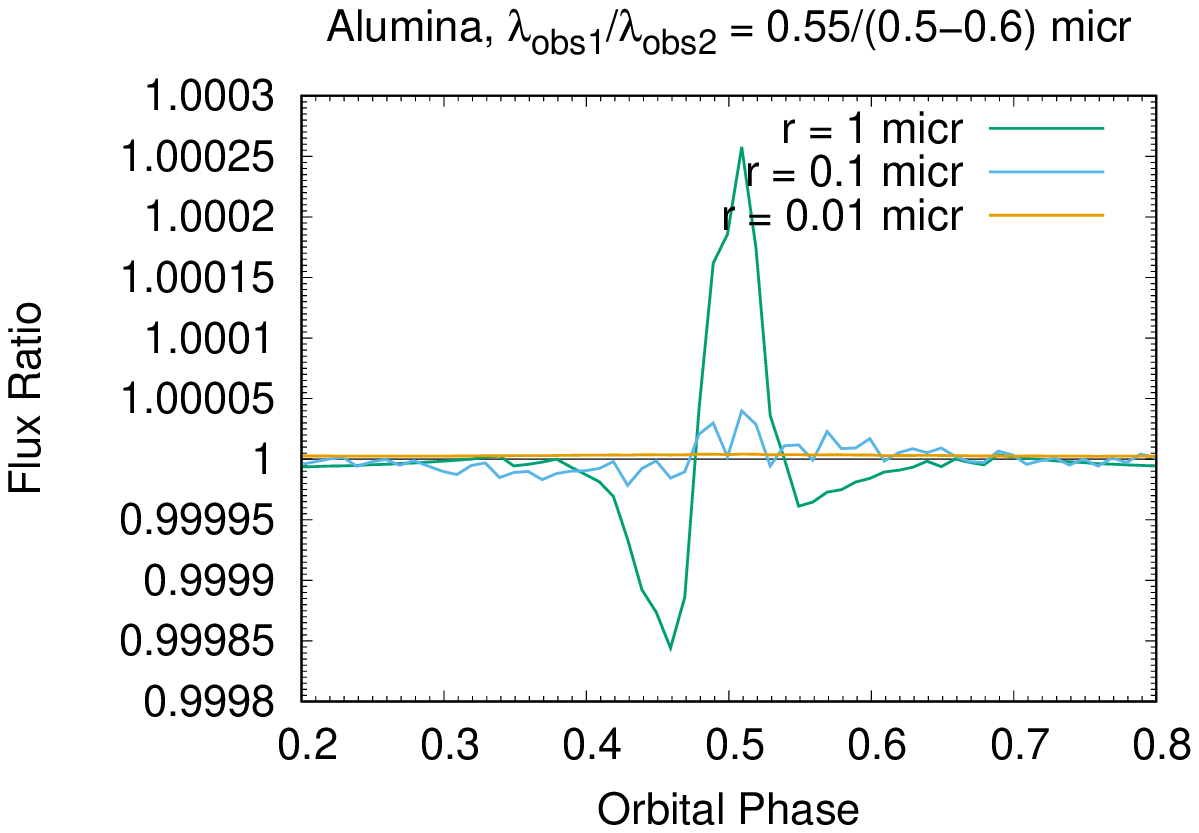}} 
\caption{Model light curves of the disintegrating exoplanet Kepler-1520b in a grazing, non-transiting regime with forward scattering only. Predicted photometry for different particle sizes of alumina, calculated for the broad-band \textit{Ariel} channel, called "VISPhot" (left-hand panel). Flux ratios between the light curves calculated for the idealized and real broad-band "VISPhot" channel are also shown (right-hand panel).}
\label{broad}    
\end{figure*}

The last factor, which we investigated is the broad-band nature of the photometry. The light-curve models presented in Section \ref{grazingmod} were calculated for a single wavelength, i.e., for the central wavelength of each \textit{Ariel} passband, however the real \textit{Ariel} observations will be taken as broad-band observations. For example, we can mention that the \textit{Ariel} channel called "VISPhot" spans from 0.50 to 0.60 micron. To see the effect of this factor on the light curves we again selected the models, which were calculated for alumina on 0.55 micron and at $i = 75^{\circ}$ without white noise (see Fig. \ref{aluminafull}, top left-hand panel), and recalculated these models for the above mentioned broad-band \textit{Ariel} "VISPhot" channel by integrating the wavelengths from 0.50 to 0.60 micron. The recalculated model light curves are depicted on the left-hand panel of Fig. \ref{broad}. For better effect visibility, flux ratios between the original and the recalculated light curves are also shown on the right-hand panel of the same figure.

As we can see, the effect of this factor is similar as in the case of the orbit inclination angle. The most affected light curve is the model calculated for 1-micron grains, where the scattering amplitude is reduced significantly in comparison with the original model calculated for the central wavelength of the passband. In comparison with the original models, the light curves calculated for 0.1- and 0.01-micron grains are relatively unchanged. Thus, we can again conclude that although the 1-micron model is theoretically the easiest to measure with \textit{Ariel}, on the other hand, the light-curve model calculated for 1-micron grains is the most sensitive to different factors. Therefore, the real detectability of a Kepler-1520b-like disintegrating planet in a grazing, non-transiting regime will strongly depend on this, and similar factors not investigated here (e.g., mixing the grain sizes in the dust-tail, red noise, etc.).  

\section{Conclusions}
\label{conc}

In our case study we took the disintegrating exoplanet Kepler-1520b and changed the orbital properties of the system to get a grazing, non-transiting orbit scenario, where the planet is just non-transiting, but part of the dust-tail is still transiting, and investigated, how different particle radii, species, observational channels, and other factors affect the amplitude of the forward-scattering peak, and the detectability of the scattering event. We analyzed the following particle radii and species: 0.01-, 0.1-, and 1-micron grains of alumina, enstatite, forsterite, olivine with 50 \% magnesium, pyroxene with 40 \% magnesium, and iron. Using the {\tt{Shellspec}} code we calculated several synthetic light-curve models on the wavelengths of the planned \textit{Ariel} space observatory, i.e., on 0.55, 0.70, 0.95, 1.25, and 1.65 microns. Our most important conclusions are the followings. 

(1) There are no significant differences among the selected species. The models are very similar from this viewpoint. 

(2) The most dominant is the 1-micron model, these particles generate forward-scattering peaks with the largest amplitude. This model is narrow and the duration of the scattering event is from 0.2 to 0.4 in units of phase. The scattering amplitude is affected by the observational channel only slightly, therefore 1-micron grains could be the most detectable material at any of the \textit{Ariel} wavelengths. On the other hand, the 1-micron model is the most sensitive to different factors. A small change in the orbit inclination angle, in the orbit radius, or the broad-band nature of the photometry can cause significant decrease in the scattering amplitude, therefore the real detectability will strongly depend on these, and similar factors not investigated here (e.g., mixing the grain sizes in the dust-tail, red noise, etc.). 

(3) A change in the observational channel affects mainly the scattering amplitude of 0.1-micron grains. This model becomes less dominant with the increasing wavelength, therefore observations on 0.55 micron is the best way to detect such a disintegrating planet. In this case the orbital inclination, the orbit radius, the broad-band nature of the photometry, or similar factors do not affect the scattering amplitude significantly. The duration of the scattering event is longer than in the previous case with wider wings near the continuum.

(4) 0.01-micron grains generate long and very small forward-scattering amplitude, which is below the detection limit.

(5) Furthermore, we investigated the impact of white noise on the detectability. In the case of Kepler-1520b in a grazing, non-transiting scenario with the typical particle size of 0.1 micron, observed on 0.55 micron, the 30-min integrated noise in the data should be $N_\mathrm{1/2h} \leq 10^{-5}$ ($\leq 10-90$ ppm) for a convincing detection. Roughly 50\% of the \textit{Ariel} core survey targets might be expected to have sufficient signal-to-noise ratio, however this optimistic number will be reduced by other factors described above. 

Based on our results we can assume that forward scattering generated by 0.1-, and 1-micron grains, evaporating from disintegrating exoplanets, creating a comet-like tail, will be detectable by \textit{Ariel} and will be possible to investigate not only transiting, but also grazing, non-transiting exoplanets based on the forward scattering. From the viewpoint of such objects the big advantage of \textit{Ariel} will be the possibility of multiwavelength observations.

\begin{acknowledgements}
I thank Prof. Gy. M. Szab\'{o} for the technical assistance, comments, and discussions. I also thank the anonymous reviewers for helpful comments and corrections. This work was supported by the Hungarian NKFI grant No. K-119517 and the GINOP grant No. 2.3.2-15-2016-00003 of the Hungarian National Research, Development and Innovation Office, by the City of Szombathely under agreement No. 67.177-21/2016, and by the VEGA grant of the Slovak Academy of Sciences No. 2/0031/18.
\end{acknowledgements}

%
%



\end{document}